\documentclass[fleqn]{emulateapj}
\usepackage{graphicx}	% Including figure files
\usepackage{amsmath}	% Advanced maths commands
\usepackage{amssymb}	% Extra maths symbols
\usepackage{bm}		% Bold maths symbols, including upright Greek
\usepackage[hidelinks=true]{hyperref}
\hypersetup{
  colorlinks   = true, %Colours links instead of ugly boxes
  urlcolor     = blue, %Colour for external hyperlinks
  linkcolor    = blue, %Colour of internal links
  citecolor   = blue %Colour of citations
}

\usepackage[T1]{fontenc}
\usepackage{newtxtext,newtxmath}
\shorttitle{Cluster Red Sequence Luminosity Function at $1.0 < z < 1.3$}
\shorttitle{Chan et al.}

\begin{document}

%%%%%%%%%%%%%%%%%%% TITLE PAGE %%%%%%%%%%%%%%%%%%%
% Title of the paper, and the short title which is used in the headers.
% Keep the title short and informative.
\title{The Rest-frame $H$-band Luminosity Function of Red Sequence Galaxies in Clusters at $1.0 < z < 1.3$}  

%\correspondingauthor{Jeffrey C.C. Chan}   (Can you add the corresponding author for me? Thanks!)
\email{E-mail: jchan@ucr.edu}

% The list of authors, and the short list which is used in the headers.
% If you need two or more lines of authors, add an extra line using \newauthor
\author{Jeffrey C.C. Chan$^{1}$, Gillian Wilson$^{1}$,  Gregory Rudnick$^{2}$,  Adam Muzzin$^{3}$,  Michael Balogh$^{4}$, Julie Nantais$^{5}$, \\ Remco F. J. van der Burg$^{6}$, Pierluigi Cerulo$^{7}$,  Andrea Biviano$^{8}$, Michael C. Cooper$^{9}$,  Ricardo Demarco$^{7}$, Ben Forrest$^{1}$, \\ Chris Lidman$^{10}$, Allison Noble$^{11}$, Lyndsay Old$^{12}$, Irene Pintos-Castro$^{13}$, Andrew M. M. Reeves$^{4}$,  Kristi A. Webb$^{4}$, \\  Howard K.C. Yee$^{13}$,   Mohamed H. Abdullah$^{1,14}$, Gabriella De Lucia$^{8}$,  Danilo Marchesini$^{15}$, Sean L. McGee$^{16}$, \\ Mauro Stefanon$^{17}$ and Dennis Zaritsky$^{18}$}
%\\
% List of institutions
%Affiliations are listed at the end of the paper\\
\affil{$^{1}$Department of Physics \& Astronomy, University of California, Riverside, 900 University Avenue, Riverside, CA 92521, USA \\
$^{2}$Department of Physics and Astronomy, The University of Kansas, Malott room 1082, 1251 Wescoe Hall Drive, Lawrence, KS 66045, USA\\
$^{3}$Department of Physics and Astronomy, York University, 4700 Keele Street, Toronto, Ontario, ON MJ3 1P3, Canada\\
$^{4}$Department of Physics and Astronomy, University of Waterloo, Waterloo, Ontario, N2L 3G1, Canada\\
$^{5}$Departamento de Ciencias F\'isicas, Universidad Andres Bello, Fernandez Concha 700, Las Condes 7591538, Santiago, Regi\'on Metropolitana, Chile\\
$^{6}$European Southern Observatory, Karl-Schwarzschild-Str. 2, 85748, Garching, Germany\\
%$^{7}$Department of Astronomy, Universidad de Concepci\'{o}n, Casilla 160-C, Concepci\'{o}n, Chile
$^{7}$Departamento de Astronom\'ia, Facultad de Ciencias F\'isicas y Matem\'aticas, Universidad de Concepci\'on, Concepci\'on, Chile \\
$^{8}$INAF-Osservatorio Astronomico di Trieste, via G. B. Tiepolo 11, 34143, Trieste, Italy\\
$^{9}$Department of Physics and Astronomy, University of California, Irvine, 4129 Frederick Reines Hall, Irvine, CA 92697, USA\\
$^{10}$The Research School of Astronomy and Astrophysics, Australian National University, ACT 2601, Australia\\
$^{11}$MIT Kavli Institute for Astrophysics and Space Research, 70 Vassar St, Cambridge, MA 02109, USA\\
$^{12}$European Space Agency (ESA), European Space Astronomy Centre (ESAC), E-28691 Villanueva de la Ca\~nada, Madrid, Spain\\
$^{13}$Department of Astronomy and Astrophysics, University of Toronto 50 St. George Street, Toronto, Ontario, M5S 3H4, Canada\\
$^{14}$Department of Astronomy, National Research Institute of Astronomy and Geophysics, 11421 Helwan, Egypt\\
$^{15}$Physics and Astronomy Department, Tufts University, Robinson Hall, Room 257, Medford, MA 02155, USA\\
$^{16}$School of Physics and Astronomy, University of Birmingham, Edgbaston, Birmingham B15 2TT, England\\
$^{17}$Leiden Observatory, Leiden University, NL-2300 RA Leiden, Netherlands\\
$^{18}$Steward Observatory and Department of Astronomy, University of Arizona, Tucson, AZ, 85719}

% These dates will be filled out by the publisher
%\date{Accepted 2100 December 3. Received 2050 December 3; in original form 2017 May 5}
%\thanks{E-mail: jchan@ucr.edu}
% Enter the current year, for the copyright statements etc.
%\pubyear{2017}
%

%I dunno what this is but this is the fix for the pdfstartlink error.
%\usepackage{etoolbox}
%\makeatletter
%\patchcmd\@combinedblfloats{\box\@outputbox}{\unvbox\@outputbox}{}{%
%   \errmessage{\noexpand\@combinedblfloats could not be patched}%
%}%
% \makeatother
%\hypersetup{draft}
% Don't change these lines
%\begin{document}
%\label{firstpage}
%\pagerange{\pageref{firstpage}--\pageref{lastpage}}
%\maketitle

% Abstract of the paper  needs to be rewrote                         (%CHECKED_REFV2)
%or $z\sim1.1$
\begin{abstract}
We present results on the rest-frame $H$-band luminosity functions (LF) of red sequence galaxies in seven clusters at $1.0 < z < 1.3$ from the Gemini Observations of Galaxies in Rich Early Environments Survey (GOGREEN).  Using deep GMOS-$z'$ and IRAC $3.6 \mu$m imaging, we identify red sequence galaxies and measure their LFs down to $M_{H} \sim M_{H}^{*} + (2.0 - 3.0)$.  By stacking the entire sample, we derive a shallow faint end slope of $ \alpha \sim -0.35^{+0.15}_{-0.15} $ and $ M_{H}^{*} \sim -23.52^{+0.15}_{-0.17} $, suggesting that there is a deficit of faint red sequence galaxies in clusters at high redshift.
By comparing the stacked red sequence LF of our sample with a sample of clusters at $z\sim0.6$, we find an evolution in the faint end of the red sequence over the $\sim2.6$ Gyr between the two samples, with the mean faint end red sequence luminosity growing by more than a factor of two.  The faint-to-luminous ratio of our sample ($0.78^{+0.19}_{-0.15}$) is consistent with the trend of decreasing ratio with increasing redshift as proposed in previous studies. A comparison with the field shows that the faint-to-luminous ratios in clusters are consistent with the field at $z\sim1.15$ and exhibit a stronger redshift dependence.  Our results support the picture that the build up of the faint red sequence galaxies occurs gradually over time and suggest that faint cluster galaxies, similar to bright cluster galaxies, experience the quenching effect induced by environment already at $z \sim 1.15$.
\end{abstract}

%Even with such a narrow redshift range, the shape of the red sequence LFs vary by a substantial amount between clusters. 
%which suggests that there is a deficient of red galaxies

\defcitealias{Rudnicketal2009}{R09}

% Select between one and six entries from the list of approved keywords.
% Don't make up new ones.

%%%%%%%%%%%%%%%%%%%%%%%%%%%%%%%%%%%%%%%%%%%%%%%%%%

%%%%%%%%%%%%%%%%% BODY OF PAPER %%%%%%%%%%%%%%%%%%

% The MNRAS class isn't designed to include a table of contents, but for this document one is useful.
% I therefore have to do some kludging to make it work without masses of blank space.
%\begingroup
%\let\clearpage\relax
%\tableofcontents
%\endgroup
%\newpage

\section{Introduction}                                                                  %CHECKED_REFV2
\label{sec:Introduction}
In the local universe, the galaxy population in the high density environment comprises mainly red, passive galaxies, as reflected by their higher quiescent fraction at fixed stellar mass than in the field \citep[e.g.][]{SandageVisvanathan1978, Baloghetal2004, Baldryetal2006, Wetzeletal2012}.  The red galaxies in the highest-density environment, i.e. galaxy clusters, have mostly early-type morphology and are mainly composed of old stellar populations \citep[e.g.][]{Dressler1980, KodamaArimoto1997, Thomasetal2005, Trageretal2008, Thomasetal2010}. They reside in a well-defined region of the color-magnitude space, known as the red sequence \citep[e.g.][]{Boweretal1992,Boweretal1998, Kodamaetal1998}.  Understanding how these red sequence galaxies form and evolve and the physical processes involved remains as one of the major goals in extragalactic astronomy. %TinsleyGunn1976

Over the last decade, much effort has been made in determining the evolution of the red sequence galaxy population in clusters out to intermediate and high redshift.  One widely used method to trace their evolution is the cluster galaxy luminosity function (LF), which measures the number of galaxies per luminosity interval.  This direct and powerful statistical tool encodes information about the star formation and mass assembly history of the galaxies, hence it can provide strong constraints for models of galaxy formation and evolution.  For example, previous studies have shown that the evolution of bright galaxies in clusters is consistent with passive evolution through studying the bright end of the red sequence cluster LF or the total cluster LF out to $z\sim1.5$ \citep[e.g.][]{Ellisetal1997, DeProprisetal1999, DeProprisetal2007, DeProprisetal2013, Linetal2006, Andreonetal2008, Muzzinetal2007, Muzzinetal2008, Rudnicketal2009, Strazzulloetal2010, Manconeetal2012}.

The extent of the evolution of the faint red sequence population, however, is still under debate.  As opposed to local clusters that exhibit a flat faint end, or even an upturn at the faint end of their red sequence LFs \citep[e.g.][]{Popessoetal2006, Agullietal2014, Morettietal2015, Lanetal2016}, various studies have revealed that clusters at intermediate and high redshifts show a continual decrease in fraction of the faint red sequence population with redshift, which indicates a gradual build up of the faint red sequence population over time since $z  \sim 1.5$ \citep[e.g.][]{Dressleretal1997, Smailetal1998, Kodamaetal2004, DeLuciaetal2004, DeLuciaetal2007, Tanakaetal2007, Gilbanketal2008, Rudnicketal2009, Stottetal2009, Rudnicketal2012, Martinetetal2015, Zentenoetal2016, Zhangetal2017, Sarronetal2018}.  This is also supported by findings that cluster galaxies on the high mass end of the red sequence are on average older than the low mass end \citep[e.g.][]{Nelanetal2005, SanchezBlazquezetal2009, Demarcoetal2010b, Smithetal2012}.  Contrary to the abovementioned studies, a number of studies have reported that there is little or no evolution in the faint end of red sequence cluster LF up to $z \sim1.5$ \citep[e.g.][]{Andreon2006, Crawfordetal2009, DeProprisetal2013, DeProprisetal2015, Andreonetal2014, Ceruloetal2016}, which in turn suggests an early formation of the faint end similar to the bright red sequence galaxies.  \citet{DeProprisetal2013} proposed that the discrepancy is primarily caused by surface brightness selection effects, which lowers the detectability of faint galaxies at high redshift.  Nevertheless, a recent study by \citet{Martinetetal2017} extensively investigated the effect of surface brightness dimming with $16$ CLASH clusters in the redshift range of $0.2 < z < 0.6$.  They concluded that surface brightness dimming alone could not explain the observed redshift evolution of the faint end.  Other possible explanations of the discrepancy invoke the radial and mass dependence of the faint red sequence population, both of which are also debated among local cluster LF studies \citep[see, e.g.][]{Popessoetal2006, Barkhouseetal2007, Lanetal2016}.  While there may be a (weak) dependence of red sequence LF on cluster mass (or cluster properties that are mass proxies) at intermediate redshift \citep[e.g.][]{DeLuciaetal2007,Muzzinetal2007, Rudnicketal2009,Martinetetal2015}, it remains unclear whether this effect exists at higher redshift.  It is also possible that the disagreements in the literature are driven by the large cluster-to-cluster variations, sample selections or the methods used to derive the LF, as observed in most of the abovementioned works.
%in turn suggests that the faint end is already in place as early as $z\sim1.5$
%are known to be responsible for

Resolving the faint end evolution is a crucial step to disentangle the underlying physical processes that quench star formation in cluster galaxies.  Mechanisms that can suppress star formation can be broadly classified into those that act internally to the galaxy and often correlate with mass (`mass-quenching'), and external processes that correlate with the environment where the galaxy resides (`environment-quenching').  Examples of mass-quenching mechanisms include feedback from supernovae, stellar winds \citep[for low-mass galaxies, e.g.][]{DekelSilk1986, Hopkinsetal2014} or active galactic nuclei (AGN) \citep[for more massive galaxies, e.g.][]{Boweretal2006, Hopkinsetal2007, Terrazasetal2016}, and heating processes that relate to the galaxy halo \citep[`halo-quenching',][]{DekelBirnboim2006, Cattaneoetal2008, Wooetal2013}.  On top of these mechanisms that are applicable to all galaxies, a galaxy can also be quenched when it enters dense environments such as galaxy groups and clusters \citep[see][for reviews]{BoselliGavazzi2006, BoselliGavazzi2014}.  As a galaxy enters a massive halo, its supply to cold gas from the cosmic web is cut off (and may also be accompanied by the stripping of hot gas in the outer parts), which results in a gradual decline of star formation as the fuel slowly runs out \citep[`strangulation' or `starvation',][]{Larson1980, Baloghetal1997}.  Quenching can also happen due to rapid stripping of the cold gas in the galaxies when it passes through the intracluster medium (ICM) \citep[`ram pressure stripping',][]{GunnGott1972} or due to gravitational interactions between galaxies to other group or cluster members, or even the parent halo \citep[`galaxy harassment', e.g.][]{Mooreetal1998}.  In the local Universe, it has been shown that the effect of mass and environmental quenching mechanisms are separable \citep[e.g.][]{Pengetal2010} and that ram-pressure stripping is able to effectively suppress star formation in cluster galaxies \citep[e.g.][]{Bosellietal2016, Fossatietal2018}.  At high redshift, the situation is more complicated.  Recent works have shown a mass dependence in the environmental quenching efficiencies at $z \gtrsim 1$ \citep{Cooperetal2010, Baloghetal2016, Kawinwanichakijetal2017, Papovichetal2018}, which suggest that the effects from both classes are no longer separable.  This points to a possible change in the dominant environmental quenching mechanism at high redshift \citep{Baloghetal2016}. A promising candidate that is supported by recent observations is the `overconsumption' model \citep{McGeeetal2014}, which suggests the gas supply in the galaxies may be exhausted by the combination of star formation and star-formation-driven outflows.  Constraining the evolution of the faint end of the cluster red sequence at high redshift is therefore important to understand the quenching mechanism and its mass dependence.  %which correspond to intermediate and low-mass galaxies, 

% that on top of the classic `strangulation' or `starvation' model, 
% which they termed as `overconsumption'.  A consequence of this `overconsumption' model is that it is more efficient on more massive galaxies, and thus will lead to a mass-dependent environmental quenching efficiency (or timescale), which is supported by recent observations \citep[e.g.][]{Baloghetal2016, Papovichetal2018}.
%Ram-pressure stripping is known to effectively and rapidly suppress star formation in cluster galaxies in the local Universe (Solanes et al. 2001; Vollmer et al. 2001; Gavazzi et al. 2010, 2013; Boselli et al. 2008, 2014b).
% These processes are complicated in nature and the exact details of their efficiency and dynamics are still poorly understood. The sit- uation is even more complicated when several of those processes are found to act together (Gavazzi et al. 2001; Vollmer et al. 2005).

In this paper we investigate the rest-frame $H$-band luminosity functions of the red sequence galaxies in seven clusters of the Gemini Observations of Galaxies in Rich Early Environments survey \citep[GOGREEN,][]{Baloghetal2017} at $1.0 < z < 1.3$.  The GOGREEN survey is an ongoing imaging and spectroscopic survey targeting 21 known overdensities at $1.0 < z < 1.5$ that are representative of the progenitors of the clusters we see today.  One of the main science goals of GOGREEN is to measure the effect of environment on low-mass galaxies. Hence, the survey aims at obtaining spectroscopic redshifts for a large number of faint galaxies down to $z' < 24.25$ and $[3.6] < 22.5$, using the Gemini Multi-Object Spectrographs (GMOS) on the Gemini North and South telescopes.  Combining all the available redshifts on these overdensities, by the end of the survey we expect to have a statistically complete sample down to stellar masses of $M_{*} \gtrsim 10^{10.3} M_\odot$ for all galaxy types.  The design of the survey and the science objectives, as well as the data reduction are described in detail in \citet{Baloghetal2017}.  %More about the redshift determination of the GOGREEN spectra will be described in a forthcoming paper.

The primary goal of this paper is to quantify the faint end of the red sequence LF and to investigate its evolution with redshift in order to shed light on the growth of the faint red sequence galaxies.  This paper is organised as follows.  A summary of the GOGREEN observations and data used in this paper are described in Section~\ref{sec:Data}.  In Section~\ref{sec:Constructing the Luminosity Function} we describe the procedure to derive membership of the galaxies, as well as the techniques used to construct the red sequence luminosity functions.  We present the luminosity functions and compare them with a low redshift sample in Section~\ref{sec:Results}.  We then compare our results with other cluster samples in the literature and the field in Section~\ref{sec:Discussion}.  In Section~\ref{sec:Conclusion} we draw our conclusions.

Throughout the paper, we assume the standard flat cosmology with $H_{0} = 70$~km~s$^{-1}$~Mpc$^{-1}$, $\Omega_{\Lambda} = 0.7 $ and $\Omega_{m} = 0.3$.  Magnitudes quoted are in the AB system \citep{OkeGunn1983}.

%%%%%%%%%%%%%%%%%%%%%%%%%%%%%%%%%%%%%%%%%%%
%%%%%%%%%       GG Survey & DATA   %%%%%%%%%%%%%%%%%%%%%
%%%%%%%%%%%%%%%%%%%%%%%%%%%%%%%%%%%%%%%%%%%
\section{Sample and Data}
\label{sec:Data}
  
\subsection{The GOGREEN Survey and Observations}                                                              %CHECKED_REFV2
\label{subsec:The GOGREEN Survey and Observations}
The cluster sample used in this paper is a subsample of the clusters observed in the GOGREEN survey \citep{Baloghetal2017}.  The full GOGREEN sample consists of three spectroscopically confirmed clusters from the South Pole Telescope (SPT) survey \citep{Brodwinetal2010, Foleyetal2011, Stalderetal2013}, nine clusters from the Spitzer Adaptation of the Red-sequence Cluster Survey \citep[SpARCS,][]{Wilsonetal2009, Muzzinetal2009, Demarcoetal2010a}, of which five were followed up extensively by the Gemini Cluster Astrophysics Spectroscopic Survey \citep[GCLASS,][]{Muzzinetal2012}, and nine group candidates selected in the COSMOS and Subaru-XMM Deep Survey (SXDS) fields.

In this study we focus on seven GOGREEN clusters at $1.0 < z < 1.3$.  The properties of the clusters are summarised in Table~\ref{tab_data_summary}.  Four of the clusters (SpARCS1051, SpARCS1616, SpARCS1634, SpARCS1638) were discovered using the red-sequence technique \citep{Wilsonetal2009, Muzzinetal2009, Demarcoetal2010a}.  The remaining three clusters were discovered via the Sunyaev-Zeldovich effect signature from the SPT survey \citep{Bleemetal2015}.  These seven clusters are chosen for their available spectroscopic coverage (from GOGREEN, SpARCS, GCLASS, and the abovementioned SPT works), so that the location of their cluster red sequence can be reliably determined. In this paper we include GOGREEN redshifts for these clusters determined from the spectra taken up to semester 2018A ($\sim77\%$ project completion).

The spectroscopic information allows us to estimate the halo mass and radius of the clusters using dynamical methods. The procedure of deriving these properties will be described in detail in a forthcoming paper (Biviano et al., in prep.). In brief, using all available redshifts of these clusters, the cluster membership of the spectroscopic objects and velocity dispersions $\sigma_v$ are determined using the \texttt{Clean} algorithm \citep{Mamonetal2013} and the new \texttt{C.L.U.M.P.S.} algorithm (Munari et al., in prep.). Both algorithms identify cluster members based on their location in projected phase-space, but while the \texttt{Clean} algorithm is based on a dynamical model for the cluster, the \texttt{C.L.U.M.P.S.} algorithm is based on the location of gaps in velocity space. The cluster $M_{200}$ is derived from the derived velocity dispersion of the clusters using the $M_{200}$ -- $\sigma_{v}$ scaling relation of \citet{Evrardetal2008}. %Using the estimated $\sigma_v$, the cluster radius $R_{200}$, i.e. the radius at which the mean interior density is $200$ times the critical density, is then estimated using an iterative procedure described in \citet{Mamonetal2013}.

We found that the $M_{200}$ value of SPT0205 is a factor of $\sim3$ lower than the value obtained by the Sunyaev-Zel'dovich effect (SZE) analysis of \citet{Rueletal2014}.   One possible explanation of this is an uncorrelated large-scale structure along the line-of-sight leading to an increase in the SZE signal, especially for low-mass clusters \citep{Guptaetal2017}.  Line-of-sight structures that are dynamically unrelated to the cluster will not be selected by the spectroscopic membership procedures, thus they would not affect the velocity dispersion estimate. However, this explanation is unlikely accurate because the SZ-derived mass is similar to the mass derived from X-ray observations by \citet{Bulbuletal2019} - if anything, X-ray derived masses tend to underestimate true cluster masses \citep{Rasiaetal2012}.  An alternative explanation for the discrepancy between the dynamical and SZ mass estimates is triaxiality. \citet{Saroetal2013} have shown that the scatter in the mass estimate from a scaling relation with the velocity dispersion is $\sim150\%$ at $z \sim 1.3$, and the scatter is mostly due to triaxiality. If SPT0205 is a very elongated cluster and if it is observed with its major axis aligned on the plane of the sky, the observed line-of-sight velocity dispersion would be much lower than the spherically averaged velocity dispersions, thereby leading to a significant underestimate of the mass via the scaling relation.  Nevertheless, we have checked that using the SZ mass estimates for this cluster instead of the dynamical estimate will not change our conclusions.

To derive the LF we make use of the deep GMOS $z'$ and \textit{Spitzer} IRAC $3.6~\mu$m images of the clusters. The details of the observation and data reduction of the images are described in \citet{Baloghetal2017}.  Below we give a brief summary of the data used in this study.

The $z'$-band imaging of the clusters were obtained using GMOS-N and GMOS-S imaging mode during September to October, 2014 and March to May 2015.  The southern clusters were observed with the Hamamatsu detector of GMOS-S with a typical exposure time of 5.4ks, while the northern clusters were observed with the e2v dd detector of GMOS-N with a long exposure time of 8.9ks to compensate for the lower sensitivity of the e2v detector.  The GMOS imaging covers a field of view (FOV) of $5\farcm5 \times 5\farcm5$. The data were reduced with the Gemini \textsc{iraf} packages with an output pixel scale of $0\farcs1458$ (e2v) or $0\farcs16$/pix (Hamamatsu), and the zero-points were determined through comparing with pre-existing CFHT/MegaCAM $z'$ imaging from SpARCS and CTIO/MOSAIC-II $z'$ imaging from the SPT collaboration. The IRAC data of the clusters come from the GCLASS \citep{vanderBurgetal2013} and SERVS \citep{Mauduitetal2012}, as well as PI programmes (PI: Brodwin, programme ID 70053 and 60099).  Available IRAC data for each cluster were combined to a  $10' \times 10'$ mosaic with a pixel scale of $0\farcs2$ per pixel using USNO-B as the astrometry reference catalogue.  %, SpUDS \citep[see][]{Galametzetal2013}, as well as

Before deriving the photometric catalogues, we first register the WCS of $z'$-band images to the $3.6~\mu$m mosaics. The WCS of the $z'$ images are fine-tuned using \textsc{gaia} in the Starlink library \citep{Berryetal2013} by comparing the coordinates of unsaturated and unblended sources on the $z'$ images to the WCS calibrated $3.6~\mu$m mosaics. The $z'$ images are then resampled to the same grid as the $3.6~\mu$m mosaics using SWarp \citep{Bertinetal2002}.

\subsection{Source detection and PSF-matched Photometry}                                                        %CHECKED_REFV2
\label{subsec:Source detection and PSF-matched Photometry}
To measure the color of the galaxy accurately, one has to make sure the measured fluxes in different bands come from the same physical projected region.  We therefore PSF-match the $z'$ images to the resolution of the $3.6~\mu$m images.  For each $z'$ and $3.6~\mu$m image a characteristic PSF is created by stacking bright unsaturated stars.  The seeing of the $z'$ images, as measured from the FWHM of the PSFs, varies between $\sim0\farcs6 - 0\farcs8$ among the clusters.  The FWHM of the $3.6~\mu$m PSFs is $\sim1\farcs8$.  With these PSFs we compute the matching kernels to degrade the $z'$ images to the $3.6~\mu$m using the \textsc{photutils} package in Astropy \citep{AstropyCollaborationetal2013}.  We check that the ratios of the growth curves of the convolved $z'$ PSF fractional encircled energy to the $3.6~\mu$m PSF deviate by $<1\%$ from unity.

Source detection and PSF-matched photometry are then performed by running SExtractor \citep{BertinArnouts1996} in dual image mode.  Although the IRAC channel is the redder band, its large-FWHM point spread function (PSF) complicates source detection due to source blending issues.  Hence here we use the unconvolved $z'$-band image as the detection band.  SExtractor is set to detect sources which have three adjacent pixels that are $\geq 1.5\sigma$ relative to local background.  Spurious detections at the boundary of the images and those at regions that have variable background due to presence of saturated bright stars \citep[see Figure 1 in][]{Baloghetal2017} are removed from the catalogue.

We use aperture magnitudes ($2''$ in diameter) from the PSF-matched $z'$ images and $3.6~\mu$m images for $z'-[3.6]$ color measurements. For galaxy total magnitudes, instead of using the heavily blended $3.6~\mu$m photometry we compute a pseudo-total $3.6~\mu$m magnitude using the abovementioned $z'-[3.6]$ color, the $z'$-band {\tt MAG\_AUTO} measurement from the unconvolved $z'$-band image and an aperture correction.  The Kron-like {\tt MAG\_AUTO} measures the flux within an area that is $2.5$ times the Kron radius \citep{Kron1980}, which is determined by the first moment of the source light profile.  It is known that {\tt MAG\_AUTO} misses a small fraction of the source flux ($\sim5\%$), especially for faint sources for which the integrated area is shrunk to its minimum allowable limit (which is set to the SExtractor default $R_{\rm{min}} = 3.5$).  To correct for this, we compute an aperture correction following the method described in \citet{Labbeetal2003} and \citet{Rudnicketal2009, Rudnicketal2012}.  We first derive the $z'$-band growth curve of stars in each cluster by stacking bright unsaturated stars in the unconvolved $z'$-band images out to $\sim7\farcs5$.  The correction needed for each galaxy is then computed by comparing its {\tt MAG\_AUTO} aperture area with the growth curve.  The median value of the correction for bright galaxies ($18.5 < [3.6] < 20.0$) is $\sim -0.03$ mag, while for faint galaxies ($22.0 < [3.6] < 23.5$) the median correction increases to $\sim-0.10$ mag.  Note that this is only a first-order correction as it assumes the objects are point sources.
%Note that this is only a first-order correction as it assumes the objects are point sources, but since we are interested in red galaxies at $z\sim1$ in this paper this is adequate to our needs.    %, an exact correction would require modeling the intrinsic profile of every galaxy. 

All magnitudes are corrected for galactic extinction using the dust map from \citet{Schlegeletal1998}, and $E(B-V)$ values from \citet{Schlaflyetal2011} and those we computed with the filter responses.  Stars are identified in the $z'$ band using the SExtractor stellarity parameter ({\tt class\_star} $\geq 0.99$) and a color cut ($z' - [3.6] < -0.14$) and are flagged in the catalogue.

To measure the completeness limit of the catalogues, we inject simulated galaxies (hereafter SGs) into the unconvolved $z'$-band images and attempt to recover them using the same SExtractor setup.  For each image, we inject 15000 SGs (10 at a time) with surface brightness profiles described by a S\'ersic profile \citep{Sersic1968}, convolved with the $z'$-band PSF.  The SGs are uniformly distributed within a total magnitude range of $20.0 < z' < 27.5$ and have similar structural parameter distributions ($n, R_e, q$) as observed galaxies at $z\sim1$, taken from \citet{vanderWeletal2014}.  The SGs are distributed randomly in image regions that are not masked by the segmentation map from SExtractor, so that the centroids of the SGs do not directly overlap with existing sources.  The recovery rate of these SGs by SExtractor gives an empirical measure of the completeness of our catalogues.  We take the magnitude that corresponds to a $90\%$ recovery rate as the completeness limit.

We also measure the formal $5\sigma$ depth of the $3.6~\mu$m images using the procedure of the empty aperture simulation described in \citet{Labbeetal2003}.  We randomly drop 1000 non-overlapping circular apertures on the image regions where no object resides. The standard deviation of the measured fluxes of these apertures gives an empirical estimation of the uncertainty in the sky level.  Using various aperture sizes, we derive a relation between aperture sizes and the measured uncertainties.

The catalogue completeness limits and the formal $5\sigma$ limits of our $2''$ aperture magnitudes computed with the relation are listed in Table~\ref{tab_data_summary}.  We use both of the limits to determine the magnitude limit for deriving the LFs (see Section~\ref{subsec:Red sequence selection} for details).

%===== GG imaging table  Biviano GOGREEN M200 and sigma Now 1DP ==============                      %CHECKED_F2_REFV2
\begin{table*}
  \caption{Summary of the imaging of the GOGREEN clusters used in this study in order of redshift.}
  \centering
  \label{tab_data_summary}
  \begin{tabular}{llcccccccc}
  \hline
  \hline
 Full Name & Name &  Redshift & $\sigma_v$\textsuperscript{a} & $M_{200}$\textsuperscript &  $R_{200}$\textsuperscript & Filter &   $[3.6]_{\rm{lim}}$\textsuperscript{b}  & Comp. limit\textsuperscript{c}  & Mag limit\textsuperscript{d} \\
           &                     &             & (kms$^{-1}$)      &   ($10^{14} M_{\sun}$)   &  (Mpc) &      &   ($5\sigma$, AB)            &      (AB)    &  (AB)    \\
  \hline
%  SpARCS J1051+5818   &  SpARCS1051  & $1.035$ &  $609 \pm 34$  &   $1.4^{+0.3}_{-0.2}$  &  $0.8 \pm 0.1$  &GN /$z'$, $[3.6]$ &   24.48  &  25.2   &  -20.20 \\ [0.5ex]   %v2
  SpARCS J1051+5818 &  SpARCS1051 & $1.035$ &  ${689 \pm 36}$    &   ${2.1^{+0.3}_{-0.3}}$  &  ${0.9 \pm 0.1}$  &GN /$z'$, $[3.6]$ &   24.48  &  ${25.1}$   &  ${-20.20}$ \\ [0.5ex]      %v3 - v528 mag limit - completeness v5
%  SPT-CL J0546--5345  &  SPT0546         & $1.067$ &  $965 \pm 69$  &   $5.5^{+1.3}_{-1.1}$  &  $1.2 \pm 0.1$  &GS /$z'$, $[3.6]$ &    24.12  & 24.8   & -20.37 \\ [0.5ex]   %v2
  SPT-CL J0546--5345  &  SPT0546         & $1.067$ &  ${1016 \pm 71}$  &   ${6.5^{+1.4}_{-1.3}}$  &  ${1.2 \pm 0.1}$  &GS /$z'$, $[3.6]$ &    24.12  & ${24.7}$   & ${-20.51}$ \\ [0.5ex]       %v3  - v528 mag limit - completeness v5
%  SPT-CL J2106--5844  &  SPT2106         & $1.132$ &  $1216 \pm 268$  &   $10.6^{+8.6}_{-5.6}$   &  $1.4 \pm 0.3$   &GS /$z'$, $[3.6]$ &    23.68  & 25.1   &  -20.55 \\ [0.5ex]  %v2
  SPT-CL J2106--5844  &  SPT2106         & $1.132$ &  ${1068 \pm 90}$  &   ${7.2^{+2.0}_{-1.7}}$   &  ${1.2 \pm 0.1}$   &GS /$z'$, $[3.6]$ &    23.68  & ${25.0}$  &  ${-20.74}$ \\ [0.5ex]    %v3 - v528 mag limit - completeness v5
%  SpARCS J1616+5545   &  SpARCS1616  & $1.156$ &  $700 \pm 39$   &   $2.0^{+0.4}_{-0.3}$  &  $0.8 \pm 0.1$  &GN /$z'$, $[3.6]$ &    24.46  & 25.1   &  -20.81 \\ [0.5ex]   %v2
  SpARCS J1616+5545   &  SpARCS1616  & $1.156$ &  ${767 \pm 38}$   &   ${2.7^{+0.4}_{-0.4}}$  &  ${0.9 \pm 0.1}$  &GN /$z'$, $[3.6]$ &    24.46  & ${25.0}$   &  ${-20.93}$ \\ [0.5ex]    %v3 - v528 mag limit - completeness v5
%  SpARCS J1634+4021   &  SpARCS1634  & $1.177$ &  $747 \pm 43$   &   $2.4^{+0.4}_{-0.4}$  &  $0.9 \pm 0.1$  &GN /$z'$, $[3.6]$ &  25.09  & 25.2  & -20.75 \\ [0.5ex]       %v2
  SpARCS J1634+4021   &  SpARCS1634  & $1.177$ &  ${715 \pm 37}$   &   ${2.1^{+0.3}_{-0.3}}$  &  ${0.9 \pm 0.1}$  &GN /$z'$, $[3.6]$ &  25.09  & ${25.1}$  & ${-20.89}$ \\ [0.5ex]        %v3  - v528 mag limit - completeness v5
%  SpARCS J1638+4038   &  SpARCS1638  & $1.196$ &  $535 \pm 32$   &   $0.9^{+0.2}_{-0.2}$  &  $0.7 \pm 0.1$  &GN /$z'$, $[3.6]$ &    25.18  & 25.3   & -20.69  \\  [0.5ex]  %v2
  SpARCS J1638+4038   &  SpARCS1638  & $1.196$ &  ${564 \pm 30}$   &   ${1.0^{+0.2}_{-0.2}}$  &  ${0.7 \pm 0.1}$  &GN /$z'$, $[3.6]$ &    25.18  & ${25.2}$  & ${-20.85}$  \\  [0.5ex]   %v3  - v528 mag limit - completeness v5
%  SPT-CL J0205--5829  &  SPT0205         & $1.320$ &  $355 \pm 34$   &   $0.3^{+0.1}_{-0.1}$  &  $0.5 \pm 0.1$  &GS /$z'$, $[3.6]$ &   23.87 & 25.2   &  -21.30 \\ [0.5ex]   %v1
%  SPT-CL J0205--5829  &  SPT0205         & $1.320$ &  $462 \pm 139$   &   $0.5^{+0.6}_{-0.4}$  &  $0.6 \pm 0.2$  &GS /$z'$, $[3.6]$ &   23.87 & 25.2   &  -21.30 \\ [0.5ex]  %v2
  SPT-CL J0205--5829    &  SPT0205          & $1.320$ &  ${678 \pm 57}$  &   ${1.7^{+0.5}_{-0.4}}$  &  ${0.7 \pm 0.1}$  &GS /$z'$, $[3.6]$ &   23.87 & ${25.1}$   &  ${-21.53}$ \\ [0.5ex]       %v3 - v528 mag limit - completeness v5
    \hline
  \multicolumn{10}{p{.925\textwidth}}{\textsuperscript{a} The velocity dispersions are measured using our dynamical analysis. See Section~\ref{subsec:The GOGREEN Survey and Observations} for details.} \\
  \multicolumn{10}{p{.925\textwidth}}{\textsuperscript{b} The quoted $5\sigma$ limits are for 2$\farcs$0 aperture magnitudes.} \\
  \multicolumn{10}{p{.925\textwidth}}{\textsuperscript{c} The $90\%$ completeness limit of the photometric catalogues, derived from the GMOS $z'$-band images. See Section~\ref{subsec:Source detection and PSF-matched Photometry} for details.} \\ 
  \multicolumn{10}{p{.925\textwidth}}{\textsuperscript{d} The cluster absolute magnitude limits in rest-frame $H$-band, used to derive the LFs. See Section~\ref{subsec:Deriving the red sequence luminosity function} for details.}
\end{tabular}
\end{table*}
%The dispersion of SPT2106 is taken from \citet{Rueletal2014} instead (the measured gapper scale).
%======================================

%%%%%%%%%%%%%%%%%%%%%%%%%%%%%%%%%%%%%%%%%%%
%%%%%%%%%      Method and analysis     %%%%%%%%%%%%%%%%%%%%                                     %ALL CHECKED
%%%%%%%%%%%%%%%%%%%%%%%%%%%%%%%%%%%%%%%%%%%
\section{Constructing the Luminosity Function}                %CHECKED_REFV2
\label{sec:Constructing the Luminosity Function}
In this section we describe the technique used to construct the red sequence luminosity functions (LFs) for the GOGREEN sample.  The spatial extent of the cluster LFs are limited by the FOV of our GMOS imaging data.  After excluding regions with lower S/N, such as the image boundaries and the regions affected by vignetting, the GMOS $z'$ images allow us to measure LF for all seven clusters up to a maximum physical radius of $\sim1 $ Mpc from the cluster center before losing area coverage. This is larger than $R_{200}$ for the five lower-mass clusters in our sample (see Table~\ref{tab_data_summary}).  To facilitate comparison with the low redshift sample (see Section~\ref{subsec:Low redshift comparison sample}), in this paper we mainly present cluster LFs that are computed within a physical radius of $R \leq 0.75$ Mpc, as limited by the low redshift sample.  Hence all figures below, unless otherwise specified, are plotted with quantities within $R \leq 0.75$ Mpc.  Choosing this radius limit also has the advantage of avoiding some image artifacts in the $z'$ data, which are due to saturated bright stars that are primarily located at the outer part of the images.  We also construct LFs computed within radii of $R \leq 0.5 R_{200}$,  $R \leq 0.5$ Mpc and $R \leq 1.0$ Mpc and will discuss them where applicable.  As we will show later, our main conclusion is not sensitive to the choice of the radius limit.
% ($5\farcm5 \times 5\farcm5$).

%%%%%% Cluster membership %%%%%% 
\subsection{Cluster membership}                %CHECKED_REFV2
\label{subsec:Cluster membership}
To construct cluster luminosity functions, it is essential to separate red-sequence galaxies that are truly cluster members from foreground or background interlopers.  The ideal way is obviously to get spectroscopic redshifts for all the galaxies in the FOV and perform dynamical analysis to determine their cluster membership (see Section~\ref{subsec:The GOGREEN Survey and Observations}). However this is very expensive for the faintest galaxies. Although the deep GOGREEN spectroscopy allows us to measure redshifts down to magnitudes $[3.6] < 22.5$,  other techniques have to be employed to determine the membership of fainter galaxies or those that are not covered in the spectroscopic sample due to spatial incompleteness.

In this study we determine the membership of the galaxies using a statistical background subtraction technique demonstrated in various works \citep[e.g.][]{AragonSalamancaetal1993, Stanfordetal1998, Smailetal1998, Andreon2006, Rudnicketal2009, Rudnicketal2012}.  This technique relies on comparing galaxy number counts of the cluster catalogue with a `control' field catalogue.  Ideally, this field catalogue should have the same depth and should contain identical passbands as the cluster catalogue.  Through comparing the catalogues in observed color-magnitude space, the excess in number counts can be converted into a probability of being a cluster member.   Other works have also utilised a photometric-redshift techniques, which uses the probability distribution of photometric redshifts to select cluster members \citep[e.g.][]{DeLuciaetal2004, Pelloetal2009}.  \citet{Rudnicketal2009} (hereafter \citetalias{Rudnicketal2009}) have demonstrated that at least for red-sequence galaxies, the statistical background subtraction technique gives consistent results in comparison to those computed using accurate photometric redshifts. This technique allows us to make full use of the deep GOGREEN $z'$ and $[3.6]$ photometry here, at the same time without needing to derive photometric redshifts. The photometric redshifts will be derived in the near future after we complete acquiring the multiwavelength imaging of the GOGREEN clusters.

%%%%%% Control field catalogue %%%%%% 
\subsubsection{Control field catalogue}                %CHECKED_REFV2
\label{subsubsec:Control field catalogue}
For the `control' field sample we utilise the publicly available deep Subaru/HSC optical ($z$) and NIR imaging ($y$) data in the COSMOS field from the Hyper Suprime-Cam Subaru Strategic Program (HSC-SSP) team and the University of Hawaii (UH) \citep{Tanakaetal2017, Aiharaetal2018}, as well as the IRAC $3.6~\mu$m data from the S-COSMOS survey \citep{Sandersetal2007}.  The UltraDeep layer of the HSC-SSP survey is the only publicly available survey that reaches depths comparable to our $z'$-band data and has a large area to overcome the effects of cosmic variance.  Due to the outstanding red sensitivities of the GMOS Hamamatsu and e2v dd detectors and the transmission of the $z'$ filters, the GMOS $z'$-band has a longer effective wavelength than the HSC $z$-band.  Hence we used both the $z$ and $y$-band data of HSC-SSP to match the passband of the GMOS $z'$-band (see Appendix~\ref{Matching of filter passbands for statistical background subtraction} for a comparison of the transmission of the filter passbands).

To ensure that the photometry of the `control' field is comparable to our clusters, we have constructed our own `control' field catalogues using the same method as the clusters.  We start by PSF-matching the HSC $z$ and $y$-band {\tt deepCoadd} images of the HSC-SSP UltraDeep layer in COSMOS (Tract UD9813) \citep[see][for details on the HSC-SSP coadd images]{Boschetal2018} to the resolution of the $3.6~\mu$m data.  We then align the images and run SExtractor in dual image mode to detect sources and perform photometry, again using the unconvolved $z$-band as the detection band.  Since a single tract of the HSC imaging is split into multiple patches, SExtractor is run on individual patches; the output catalogues are then visually checked to remove spurious detections and are combined into a single master catalogue.  For each galaxy we derive an aperture correction to convert {\tt MAG\_AUTO} to a total magnitude using a stacked growth curve of bright unsaturated stars in the corresponding patch.  Patches that have depth shallower than our GMOS data or are affected by bright saturated stars and image artifacts were excluded. The final field catalogue contains $\sim450000$ galaxies and covers an area of $1.03$ deg$^{2}$.

%%%%%% Statistical subtraction %%%%%% 
\subsubsection{Membership probabilities}                %CHECKED_REFV2
\label{subsubsec:Membership probabilities}
We adopt the method used in \citet{Pimbbletetal2002} and \citetalias{Rudnicketal2009} to statistically compare the galaxy number counts between the cluster and the field sample.  In brief, for each cluster we construct the $z' - [3.6]$ vs $[3.6]$ color--magnitude diagram and the equivalent color--magnitude diagram for the field.  A color term is derived using SSP models to match the filter passbands between the cluster ($z'$) and field ($z, y$) catalogue (see Appendix~\ref{Matching of filter passbands for statistical background subtraction} for details).

The cluster galaxy population that satisfies the area selection (e.g. $R \leq 0.75$ Mpc) and the field sample are binned in color--magnitude space with bins of 0.5 mag both in color and $[3.6]$ magnitude. The number counts of the field in each bin are then scaled to the same area selection used for the cluster.  By comparing the cluster and field galaxy number counts in each bin, we can assign a cluster membership probability ($P_{\rm{memb}}$) to each galaxy in the cluster sample.  Spectroscopically confirmed cluster members as determined from dynamical analysis (see Section~\ref{subsec:The GOGREEN Survey and Observations}) are pre-assigned to have a probability of $1$.  Similarly, confirmed interlopers are pre-assigned to have a probability of $0$. The probabilities of the rest of the galaxies in each bin are then assigned as:

\begin{equation}
  \label{eqt-clus}
   P_{\rm{memb}} = 1 - \frac{ F_{A}~N_{\rm{field}} - N_{\rm{interloper}} }{ N_{\rm{cluster}} - N_{\rm{interloper}} - N_{\rm{specmemb}} }
\end{equation}

\noindent where $F_{A}$ is the scaling factor to scale the area coverage of the field to the area of the cluster in consideration.  $N_{\rm{field}}$ and $N_{\rm{cluster}}$ are the number of galaxies of the field and cluster sample in that particular bin, where $N_{\rm{interloper}}$ and $ N_{\rm{specmemb}}$ correspond to the number of spectroscopically confirmed interlopers and cluster members, respectively. From Equation~\ref{eqt-clus} one can also see that the probability will not be well defined if $F_{A}~N_{\rm{field}} > (N_{\rm{cluster}} - N_{\rm{specmemb}})$. To solve this we follow the approach of \citet{Pimbbletetal2002} to expand the color and magnitude selection used to calculate this probability by merging neighbouring bins until the resultant probability reaches $0 \leq P_{\rm{memb}} \leq 1$.

Note that the total sum of probabilities within a bin (or, equivalently, the effective number of cluster members $N_{\textrm{eff}}$) is always set by statistical background subtraction. The numbers of both confirmed members and interlopers are folded in Equation~\ref{eqt-clus}, so that the membership probability of the rest of the galaxies in the bin would be adjusted accordingly to conserve the total sum of probabilities.  %We present a test comparing the spectroscopic membership as determined from the dynamical analysis with the photometric membership probability in Appendix~\ref{app:Comparison of the spectroscopic membership with photometric membership probabilities}.

To derive the uncertainty of the background subtraction, instead of computing the probability by rescaling the entire COSMOS catalogue, we perform the same process with randomly selected regions in the catalogue with the same area as the cluster. This allows us to derive the resultant $1\sigma$ field-to-field variation within COSMOS. In addition, we derive the uncertainty in the number of galaxies of the COSMOS field sample due to cosmic variance, following the recipe in \citet{Mosteretal2011}. The two uncertainties are added in quadrature, and the combined uncertainty is then used as the uncertainty of the probabilities.

%%%%%% RS selection %%%%%%
\subsection{Red sequence selection}                %CHECKED_REFV2
\label{subsec:Red sequence selection}
The red sequence galaxies of the cluster sample are identified using $z' - [3.6]$ vs $[3.6]$ color magnitude relations (CMR).  We fit the CMR for each cluster using a fixed slope of $-0.09$.  Due to the low contrast of the cluster red sequence against interlopers, only spectroscopically confirmed galaxies that have no visually identifiable [\ion{O}{2}] emission lines are used to derive the fit.  The chosen slope of $-0.09$ is determined by fitting the CMR of SpARCS1616 and SPT0546, the two clusters that have a large number of spectroscopically confirmed galaxies, which allows us to reliably determine the slope and the zero-point simultaneously. We note that this is also the same slope of the CMR found in \citet[][]{DeLuciaetal2004, DeLuciaetal2007} and R09.  The potential red sequence galaxies are selected as galaxies within $\pm 0.25$ mag of the fitted CMRs.  Since some of the clusters have only a small number of spectroscopically confirmed members, we use a fixed magnitude selection for all the clusters. The $0.25$ mag selection corresponds to $\sim1.5-2.0\sigma$ of the intrinsic scatter of the fitted CMRs.  We verify that varying the slope by $\pm0.1$ (i.e. $0.01, -0.19$) or increasing the red sequence selection to $\pm 0.4$ mag do not change our main conclusion.

We also identify the brightest cluster galaxy (BCG) in each cluster using a simple ranking system.  Three scores are assigned to each galaxy according to a) its $[3.6]$ total magnitude (score:  3 if $[3.6] \leq 18.5$, 2: $18.5 < [3.6] \leq 19.5$, 1: $19.5 < [3.6] \leq 20$),  b) distance to the cluster centroid (3: $R \leq 0.25$~Mpc,  2: 0.25 $< R \leq 0.5$~Mpc,  1: 0.5 $< R \leq 1.0$~Mpc) and c) $z' - [3.6]$ color (3: $z'- [3.6] > $~CMR$-0.5$,  2: CMR$-1.5 < z'- [3.6] \leq $~CMR-0.5,  1: $0 \leq z'- [3.6] \leq $~CMR-1.5.  The highly-scored candidates (with a total score $\geq 4$) are then visually examined to determine the most probable BCG.  In most clusters, the BCG can be clearly identified. The candidate BCG is usually the brightest confirmed cluster member in $[3.6]$ within the uncertainties, except in SpARCS1634, where there exists one other member that is significantly off-centered ($\sim500$ kpc) and is brighter in $[3.6]$ than the assigned BCG.  We test that choosing this galaxy as the cluster BCG instead does not change our conclusion.

\begin{figure*}
  \centering
  \includegraphics[scale=1.15]{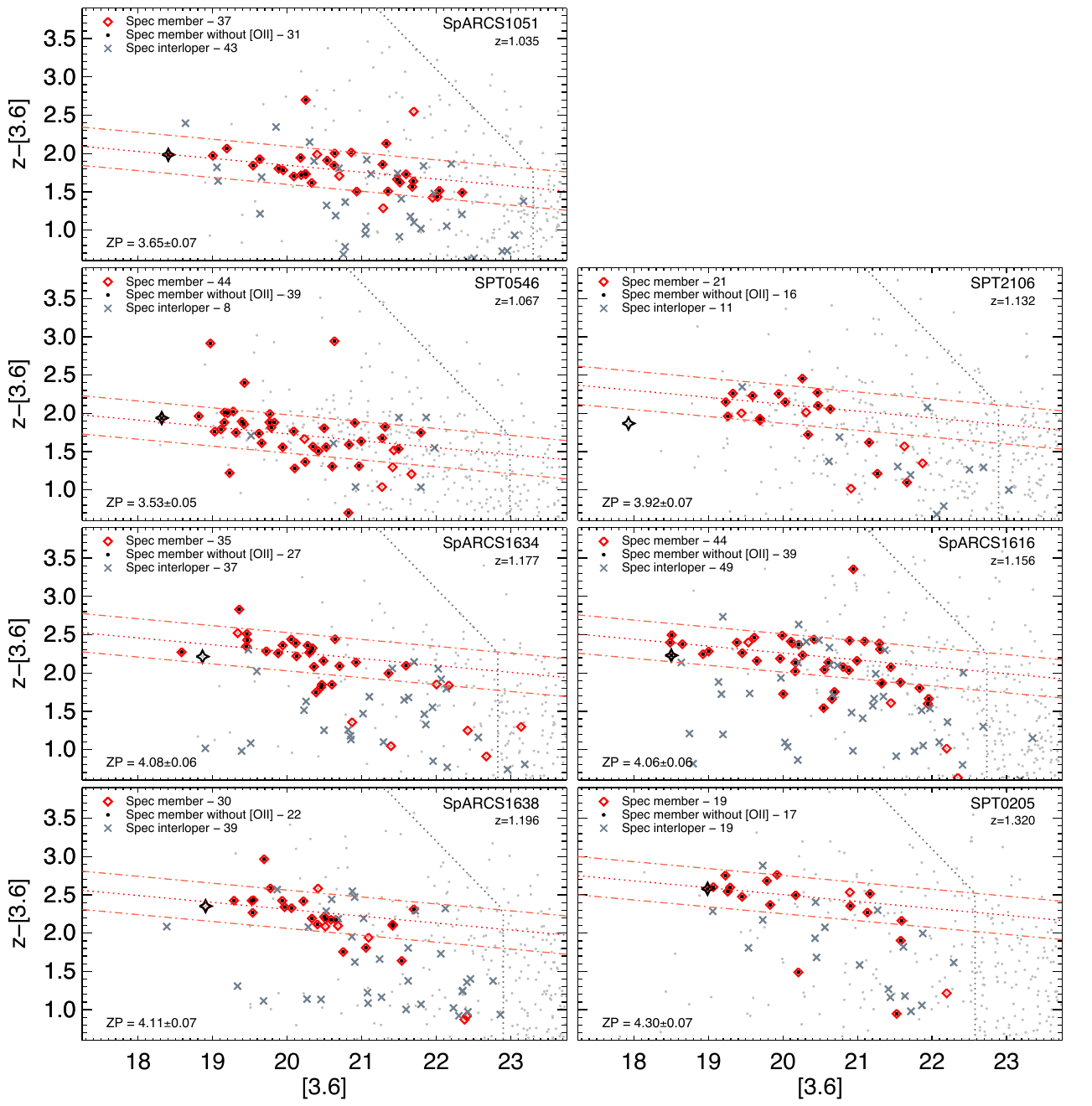}
  \caption{Color--magnitude diagram of the seven GOGREEN clusters used in this study in order of increasing redshift.  The photometry is described in Section~\ref{subsec:Source detection and PSF-matched Photometry}. The red diamonds correspond to spectroscopically confirmed cluster members.  Passive members that do not have obvious [\ion{O}{2}] emission are marked in black.  The dark grey crosses mark the interlopers, and the black star symbol corresponds to the BCG of the clusters. The red dotted line in each panel corresponds to the fitted CMR. The dotted-dashed lines in each panel correspond to $\pm0.25$ mag from the fitted relation, which is the region we used for our red sequence selection.  The dark grey dotted lines mark the magnitude limit of each cluster (see Section~\ref{subsec:Red sequence selection} for details).  The zero-point (ZP) of the CMR at $[3.6] = 0$ is given in each panel.  It can be seen that the fixed slope we adopt ($-0.09$) describes the CMR of all the clusters well.}
  \label{fig_cmd}
\end{figure*}

Figure~\ref{fig_cmd} shows the color magnitude diagram of the GOGREEN clusters and their fitted CMRs.  The zero-point (ZP) of the CMR at $[3.6] = 0$ is given in each panel.  Note that at the time of writing this paper the data acquisition for GOGREEN is still ongoing, hence some of the clusters show a lack of spectroscopic members at faint magnitudes.  The fully-completed GOGREEN spectroscopic sample will be complete down to $[3.6] < 22.5$.  The magnitude limit of each cluster is set to be the brighter magnitude between its $90\%$ completeness limit (after converting into $[3.6]$ using the red sequence color) and the $5\sigma$ limit of the $3.6~\mu$m image.  We find that for all the GOGREEN clusters the magnitude limit is set by the $90\%$ completeness limit.

%==== CMD =====================     %CHECKED_REFV2
%=====================================
%Grey crosses correspond to spectroscopically confirmed interlopers.

%%%%%% Luminosity function fits %%%%%%
\subsection{Deriving the red sequence luminosity function}            %CHECKED_REFV2
\label{subsec:Deriving the red sequence luminosity function}
At $1.0 < z < 1.3$, the $3.6~\mu$m images correspond roughly to rest-frame $H$-band.  We derive $k$-correction factors using \citet{BruzualCharlot2003} stellar population models and the software \textsc{EzGal} \citep{ManconeGonzalez2012} to convert the $[3.6]$ into rest-frame $H$-band magnitudes.  We assume a model with a formation redshift of $z_f = 3.0$, $Z=Z_\odot$ and a \citet{Chabrier2003} initial mass function (IMF).  We have checked that this model is able to reproduce the red sequence color of the clusters at their particular redshifts. The $k$-corrections range from $\sim -0.71$ to $-0.89$ depending on the redshift of the cluster.  The absolute magnitude limit of the clusters in rest-frame $H$-band is listed in Table~\ref{tab_data_summary}.

Assuming the observed cluster LF can be described by a single \citet{Schechter1976} function $\Phi (M)$, we construct the LF and derive the Schechter parameters for each cluster, including the characteristic magnitude $M^{*}$, the faint end slope $\alpha$ that characterises the power-law behaviour at magnitudes fainter than $M^{*}$, and the normalisation $\Phi_{*}$ using two different approaches:

\begin{enumerate}
 \item The binning method.  Based on the cluster membership probabilities we computed in Section~\ref{subsec:Cluster membership}, we derive 1000 Monte Carlo realizations of the red sequence sample for each cluster.  The red sequence realizations are binned in rest-frame $H$-band absolute magnitudes with a 0.5 mag bin width to the cluster magnitude limit.  The LF is then derived by taking the average of the number of galaxies of the realizations in each magnitude bin.
  
The error budget of each magnitude bin of the LF comprises the Poisson noise on the number of galaxies in the bin, computed using the recipe of \citet{Ebeling2003}, and the uncertainty of the background subtraction (i.e. from the membership probabilities).

The binned LFs are fitted with a single \citet{Schechter1976} function.  Since the Schechter parameters are highly degenerate, we follow the $\chi^2$ grid fitting approach described in  \citetalias{Rudnicketal2009}, which samples the parameter space and reduces the chance of the fit trapping in some local $\chi^2$ minima.  We start by constructing a coarse grid of $\phi_{*}, M_{H}^{*}, \alpha$.  The Poisson error of each bin is first symmetrized, the $\chi^2$ value at each grid point is evaluated, and a finer grid is then constructed using the set of parameters that give the lowest $\chi^2$ as the new centroid of the grid.  This process is iterated for two more times to derive the best-fit.  For each grid point we convert the $\chi^2$ value into a probability with $P = \exp(-\chi^2/2)$.  The $1\sigma$ uncertainty of each parameter is then determined through marginalising the other two parameters to obtain a probability distribution and taking the bounds that encloses the 16\% and 84\% of the probability distribution.

 \item The maximum likelihood estimation (MLE) method.  We also derive the LF using a parametric maximum likelihood estimator.  The standard MLE method (i.e. the STY method), first proposed by \citet{Sandageetal1979}, has been used in various LF studies.  In this paper we use a modified MLE approach to account for the cluster membership probabilities. The best fit is found by maximising the following log-likelihood function:

\begin{equation}
  \label{eqt-llf}
  \centering
   \ln \mathcal{L} = \sum_{i=1}^{N}  \left( P_{\rm{memb,i}} \times  \ln P(M_i) \right)
\end{equation}
 
\noindent where $P_{\rm{memb,i}}$ is the cluster membership probability for each galaxy described in Section~\ref{subsec:Cluster membership}, and $P(M_i)$ is the probability of observing a galaxy of absolute magnitude $M_i$ according to the \citet{Schechter1976} function:

\begin{equation}
  \label{eqt-llf2}
  \centering
  P(M_i) \propto \frac{ \Phi(M_i) }{ \int_{-\infty}^{M_{\rm{lim}}} \Phi(M) d M }
\end{equation}

\noindent The upper limit $M_{\rm{lim}}$ is set to be the magnitude limit for each cluster.  Note that strictly speaking this method also involves binning of the data as well, as $P_{\rm{memb,i}}$ is derived in binned color-magnitude space.  $P_{\rm{memb,i}}$ is incorporated into Equation~\ref{eqt-llf} in a way such that it is equivalent to running Monte Carlo realizations of the MLE derivation with the probabilities. To estimate the uncertainty of the Schechter parameters of the fit we follow the prescription described in \citet{Marchesinietal2007} to determine the error contours of $M^{*}$ and $\alpha$ from the values of the log-likelihood. The 68\%, 95\%, and 99\% confidence level are estimated by finding the ellipsoids that satisfy $\ln \mathcal{L} = \ln \mathcal{L}_{\rm{max}} - 0.5 \chi_\beta^{2}(2)$ where $\chi_\beta^{2}(2) = 2.3, 6.2, 11.8$, respectively.  To propagate the uncertainty of the membership probabilities due to field-to-field variation within COSMOS and cosmic variance, we also derive 500 Monte Carlo realizations of the red sequence probabilities and repeat the MLE fit.  The $1\sigma$ variation of the best-fit Schechter parameters from these realizations is then added in quadrature with the abovementioned uncertainty of the Schechter parameters.  In all cases, the $1\sigma$ variation of the Schechter parameters due to the uncertainty of the probabilities is much smaller than the uncertainty of the Schechter parameters of the fit.
\end{enumerate}

In both methods we exclude the BCG and galaxies brighter than the BCG when constructing the LF, as is common practice.

%%%%%% Composite red sequence LF %%%%%%
\subsection{Composite red sequence luminosity function}            %CHECKED_REFV2
\label{subsec:Composite red sequence luminosity function}
The number of galaxies in the LFs of high redshift clusters is often too low to reliably determine Schechter parameters.  Hence beside individual cluster LFs we also derive composite red sequence LF by combining the sample to measure cluster average properties.  Before stacking the LFs, a passive evolution correction is applied to the rest-frame absolute magnitudes $M_H$ to bring all clusters to the mean redshift of the sample at $\bar{z}\sim1.15$.  This correction is again computed using \citet{BruzualCharlot2003} stellar population models ($z_f = 3.0$, $Z=Z_\odot$ and a \citet{Chabrier2003} IMF).  The corrections range from $\sim -0.05$ to $0.13$ depending on the redshift of the cluster.  Similar to Section~\ref{subsec:Deriving the red sequence luminosity function}, the composite LF is also derived and fitted with both the binning approach and the MLE approach.

For the former approach, we adopt the method of \citet{Colless1989} to combine individual cluster LFs into a single composite.  The \citet{Colless1989} method combines individual LFs by renormalising the bin counts with the total number of galaxies in each LF down to a certain renormalisation magnitude limit (i.e. to convert the number counts in a particular magnitude to a fraction of the sample) and summing these renormalised counts.  Therefore the cluster LFs are normalised to the same effective richness before being combined into a single composite. The renormalisation magnitude limit has to be brighter than the magnitude limits of all the clusters being stacked, and, at the same time faint enough so that the total number of galaxies used for renormalisation is representative of the richness of the clusters \citep{Popessoetal2005}. For our GOGREEN sample the renormalisation magnitude limit is chosen to be $M_H = -21.5$.  We then fit the composite LF down to the brightest magnitude of the magnitude limits of the individual clusters.

For the MLE approach, we derive the best-fitting Schechter function using the entire red sequence sample and their corresponding cluster membership probabilities down to the same magnitude limits as the binning approach.  Similar to the individual cluster LF, we derive 500 Monte Carlo realizations of the red sequence probabilities using the uncertainty of the membership probabilities and repeat the MLE fit. The $1\sigma$ variation of the best-fit Schechter parameters from these realizations is then added in quadrature with the fitting uncertainties, and the combined uncertainty is used as the uncertainty of the Schechter parameters. We present the composite LFs derived with both approaches in Section~\ref{sec:Results}.
%We have checked that this model is able to reproduce the red sequence color of the clusters at their particular redshifts.  

%%%%%% Faint-to-luminous Ratio %%%%%%% 
\subsection{Faint-to-luminous Ratio}                                              %CHECKED_REFV2
\label{subsec:Faint-to-luminous Ratio}
Another quantity that is commonly used in previous studies to describe the luminosity distribution of red sequence galaxies, is the faint-to-luminous ratio (or dwarf-to-giant ratio).  The faint-to-luminous ratio is simply the ratio of the number of faint galaxies within a certain magnitude range to the number of those brighter than this faint population.  Essentially being a two-bin LF, the faint-to-luminous ratio is a simple quantity that is easy to compute and compare straightforwardly with other samples, without needing to assume any functional form of the underlying galaxy luminosity distribution \citep{GilbankBalogh2008}.

Here we adopt the definition of the faint-to-luminous ratio as in \citet{DeLuciaetal2004, DeLuciaetal2007} to enable a comparison with earlier works.  Luminous red sequence galaxies are defined as galaxies with rest-frame $V$-band Vega magnitude $M_{V,\rm{vega}} \leq -20$, and faint red sequence galaxies are those with $-20 < M_{V,\rm{vega}} \leq -18.2$.  We apply $k$-corrections and evolution corrections to the $z'$-band total magnitudes of our red sequence sample to convert them into rest-frame $V$-band magnitude at $z=0$.  The corrections are again computed using BC03 models, assuming $z_f = 3.0$ and $Z=Z_\odot$.  The combined corrections (not including the distance moduli) range from $\sim -1.35$ to $-1.12$ depending on the redshift of the cluster.

Similar to the LF we compute the ratio for regions within a radius of $R \leq 0.75$ Mpc.  For all clusters except SPT0205, we run 10000 random realisations of the red sequence sample using the cluster membership probabilities from Section~\ref{subsubsec:Membership probabilities}, varying also the galaxy magnitudes within their uncertainties.  Following the above definition, we then compute the faint-to-luminous ratio for each realisation, and take the median and the $1\sigma$ scatter of the distribution as the cluster faint-to-luminous ratio and its associated uncertainty.  This method is not applicable to SPT0205, as its rest-frame $V$-band depth (converted from $z'$-band) is not deep enough to compute the number of faint galaxies.   Hence for SPT0205 we first derive the rest-frame $V$-band LF and its best-fitting Schechter function using the same red sequence selection and fitting method described in Section~\ref{subsec:Deriving the red sequence luminosity function}.  The only difference is that the LF is derived in rest-frame $V$-band instead of $H$-band.  We then integrate the best fit down to $-18.2$ to extrapolate the number of faint galaxies for the faint-to-luminous ratio.  The uncertainties of the Schechter parameters are propagated to the computed ratio.  Due to the large uncertainty of the Schechter fit, the faint-to-luminous ratio computed in this way has a considerably larger uncertainty.  

We also compute the cluster-average faint-to-luminous ratio for the entire sample by integrating the best fitting Schechter function of the rest-frame $V$-band composite LF.

%%%%%% Low redshift comparison sample %%%%%%
\subsection{Low redshift comparison sample}                                    %CHECKED_REFV2
\label{subsec:Low redshift comparison sample}
As we mentioned in the introduction, one of the primary goals of this work is to investigate the evolution of the red sequence LF with redshift.  Several works have demonstrated that using different filter passbands, methods to determine cluster membership, and procedure to construct LF can affect the derived Schechter parameters to a large extent \citep[see, e.g.][]{Alshinoetal2010}. Therefore to ensure the comparison is accurate, we decide to construct our own low-redshift comparison instead of comparing our results to LF derived in previous works.

%20 optically
The low redshift comparison sample we used in this study is from the ESO Distant Cluster Survey \citep[EDisCS,][]{Whiteetal2005}, which targets optically selected cluster fields in the redshift range of $0.4 < z < 1.0$ from the Las Campanas Distant Cluster Survey \citep[LCDCS,][]{Gonzalezetal2001}.  The rest-frame $g$,$r$ and $i$-band red sequence LFs of sixteen EDisCS clusters are studied in detail in \citetalias{Rudnicketal2009}.  To avoid wavelength-dependent effects and possible biases due to the procedure used, we re-derive cluster LFs using the EDisCS photometric and spectroscopic catalogues \citep[][R09]{Whiteetal2005, Hallidayetal2004, MilvangJensenetal2008, Pelloetal2009} with filter bands that are comparable in rest-frame wavelength with our GOGREEN sample. The EDisCS photometric catalogue comprises photometry in either $B,V,I,K_s$ or $V,R,I,J,K_s$-bands from VLT/FORS2 and NTT/SOFI depending on the redshift of the cluster.  To mimic the $z' - [3.6]$ selection used for the GOGREEN sample, we identify red sequence candidates through fitting the CMR in $R-K_s$ vs $K_s$ (or $V-K_s$ vs $K_s$ if $R$-band is not available) for clusters with $z < 0.57$. For higher redshift clusters the CMR is fitted in $I-K_s$ vs $K_s$.  Among the cluster sample in \citetalias{Rudnicketal2009}, we exclude the clusters CL1354-1230 and CL1059-1253 as they have insufficient depth in the $K_s$-band image, hence we arrive at a sample of fourteen clusters.  The properties of the clusters can be found in Table~\ref{tab_data_summary_ediscs}.  We have checked that the choice of the color does not largely impact the red sequence selection. For most clusters selecting with $R-K_s$ or $I-K_s$ color gives consistent results.  %, except for those that have multiple clusters in the same FOV.  Since the cluster members of these foreground / background systems will likely bias our membership determination we also remove these clusters from the sample.  

To determine the cluster membership probabilities we follow the statistical background subtraction method outlined in Section~\ref{subsec:Cluster membership}. For the EDisCS clusters the COSMOS/UltraVISTA catalogue \citep[DR1,][]{Muzzinetal2013c} is used as the control field catalogue.  The UltraVISTA photometry are derived in a similar way as the EDisCS clusters. The large area coverage ($\sim1.8$ deg$^{2}$) and the photometric bands, including the optical ($u^{*},g^{+},r^{+},i^{+},z^{+},B_j,V_j$ + 12 medium bands) and deep NIR ($Y,J,H,K_s$) photometry make it the perfect candidate for this purpose.  Note that this is a different field sample as the one used in \citetalias{Rudnicketal2009} as we are measuring the LF in $K_s$-band.

We then apply $k$-corrections and evolution corrections to convert the $K_s$ magnitudes to rest-frame $H$-band magnitudes at the mean redshift of the selected EDisCS clusters ($\bar{z}\sim0.60$), and derive rest-frame $H$-band composite LFs following the same procedure and fitting methods described in Section~\ref{subsec:Deriving the red sequence luminosity function} and~\ref{subsec:Composite red sequence luminosity function}.  Instead of using the EDisCS catalogue completeness limit ($I \sim 24.9$) as the magnitude limit for fitting the LF, we measure $5\sigma$ magnitude limits of the $V,R,I,K_s$ bands from the uncertainties of the galaxies in the photometric catalogues and compute the corresponding magnitude limits in rest-frame $H$-band.  We found that the $K_s$-band magnitude limit is always the brightest among all the bands, thus it is used as the magnitude limit for fitting the LF.  Note that the $K_s$-band limits ($K_s \sim 21.0 - 22.3$) are also brighter than the completeness limit converted to $K_s$-band using the $I-K_s$ color of the red sequences ($\sim 1.0 - 2.5$).\\

%===== EDISCS cluster table from R09 ==============
\begin{table}
  \caption{Summary of the properties of the 14 EDisCS clusters used for comparison.}   %CHECKED_REFV2
  \centering
  \label{tab_data_summary_ediscs}
  \begin{tabular}{@{}lccccc@{}}
  \hline
  \hline
Name &  Redshift & $M_{200}$\textsuperscript{a} &  $R_{200}$\textsuperscript & Filter\textsuperscript{b}    & Mag limit\textsuperscript{c} \\
           &                &  ($10^{14} M_{\sun}$)        &  (Mpc)                                 &             &      (AB)                                \\
  \hline
 CL1216-1201 &    0.794 & $     7.6^{+  1.7}_{-  1.6}$  &  $     1.4^{+  0.1}_{-  0.1}$  &   $I,K_s$    &   -20.97 \\ [0.5ex]
% CL1354-1230 &    0.762 & $     2.0^{+  1.1}_{-  0.9}$  &  $     0.9^{+  0.1}_{-  0.2}$  &   $I,K_s$    &  -21.31  \\ [0.5ex]
 CL1054-1245 &    0.750 & $     1.0^{+  0.8}_{-  0.3}$  &  $     0.7^{+  0.2}_{-  0.1}$  &   $I,K_s$    &  -20.55  \\ [0.5ex]
 CL1040-1155 &    0.704 & $     0.6^{+  0.3}_{-  0.2}$  &  $     0.6^{+  0.1}_{-  0.1}$  &   $I,K_s$   &  -20.79  \\ [0.5ex]
 CL1054-1146 &    0.697 & $     1.6^{+  0.7}_{-  0.5}$  &  $     0.9^{+  0.1}_{-  0.1}$  &   $I,K_s$    &  -20.81  \\ [0.5ex]
 CL1227-1138 &    0.636 & $     1.5^{+  0.6}_{-  0.5}$  &  $     0.9^{+  0.1}_{-  0.1}$  &   $I,K_s$    &  -20.81  \\ [0.5ex]
  CL1353-1137 &    0.588 & $     2.4^{+  1.8}_{-  1.2}$  &  $     1.0^{+  0.2}_{-  0.2}$  &   $I,K_s$    &  -20.61  \\ [0.5ex]    
 CL1037-1243 &    0.578 & $     0.3^{+  0.2}_{-  0.1}$  &  $     0.5^{+  0.1}_{-  0.1}$  &   $I,K_s$    &  -20.63   \\ [0.5ex]    
 CL1232-1250 &    0.541 & $    10.6^{+  3.9}_{-  2.4}$  &  $     1.7^{+  0.2}_{-  0.1}$  &   $V,K_s$    &  -20.53  \\ [0.5ex]
  CL1411-1148 &    0.519 & $     3.1^{+  1.9}_{-  1.4}$  &  $     1.2^{+  0.2}_{-  0.2}$  &   $V,K_s$    &  -20.76  \\ [0.5ex]
 CL1420-1236 &    0.496 & $     0.1^{+  0.1}_{-  0.1}$  &  $     0.4^{+  0.1}_{-  0.1}$  &   $V,K_s$    &  -20.68  \\ [0.5ex]
 CL1301-1139 &    0.483 & $     2.8^{+  1.1}_{-  0.9}$  &  $     1.1^{+  0.1}_{-  0.1}$  &   $V,K_s$    &  -20.56  \\ [0.5ex]
 CL1138-1133 &    0.480 & $     3.4^{+  1.1}_{-  1.0}$  &  $     1.2^{+  0.1}_{-  0.1}$  &   $R,K_s$    &  -20.12  \\ [0.5ex]
 CL1018-1211 &    0.474 & $     1.0^{+  0.4}_{-  0.3}$  &  $     0.8^{+  0.1}_{-  0.1}$  &   $V,K_s$    &  -20.50  \\ [0.5ex]
% CL1059-1253 &    0.456 & $     1.2^{+  0.4}_{-  0.3}$  &  $     0.9^{+  0.1}_{-  0.1}$  &   $V,K_s$    &  -20.78  \\ [0.5ex]
 CL1202-1224 &    0.424 & $     1.3^{+  0.8}_{-  0.6}$  &  $     0.9^{+  0.2}_{-  0.2}$  &   $V,K_s$    &  -19.92  \\ [0.5ex]
   \hline
  \multicolumn{6}{p{.46\textwidth}}{\textsuperscript{a} The cluster $M_{200}$ is estimated using the $M_{200}$ -- $\sigma_{v}$ relation from \citet{Evrardetal2008}. The $\sigma_{v}$ are taken from the EDisCS photometric catalogues.} \\
  \multicolumn{6}{p{.46\textwidth}}{\textsuperscript{b} The bands we used to fit the cluster CMR for red sequence selection.} \\
  \multicolumn{6}{p{.46\textwidth}}{\textsuperscript{c} The cluster absolute magnitude limits in rest-frame $H$-band, used to derive the LFs. See Section~\ref{subsec:Low redshift comparison sample} for details.}
\end{tabular}
\end{table}
%============================================

%%%%%%%%%%%%%%%%%%%%%%%%%%%%%%%%%%%%%%%%%%%
%%%%%%%%%     RESULTS      %%%%%%%%%%%%%%%%%%%%%%%%%
%%%%%%%%%%%%%%%%%%%%%%%%%%%%%%%%%%%%%%%%%%%
\section{Results}                                                                          %CHECKED_REFV2
\label{sec:Results}
In this section we present the red sequence LFs of the GOGREEN clusters.  We will start by presenting the red sequence LF of individual clusters, followed by the composite LFs and the comparison with the low redshift sample.  For simplicity, the LFs are shown in galaxy number counts ($\log(N)$) in all figures.  %but note that since they are all computed within a physical radius of $R \leq 0.75$ Mpc, these are equivalent to LF in units of number density.

%%%%%% Individual LF %%%%%%% 
\subsection{Individual luminosity function}                                  %CHECKED_REFV2
\begin{figure*}
  \centering
  \includegraphics[scale=1.275]{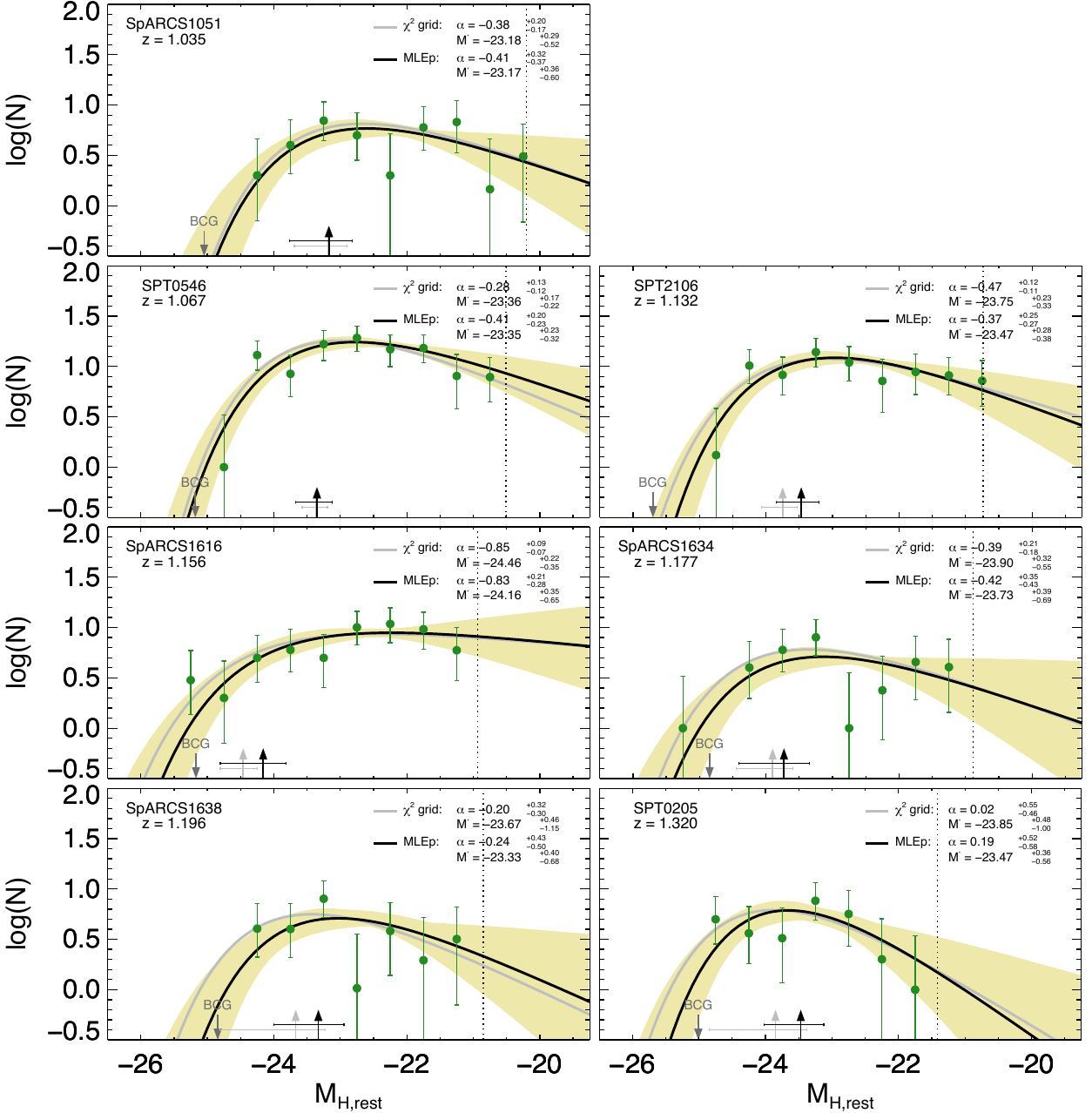}
  \caption{Rest-frame $H$-band red sequence LF of the seven GOGREEN clusters included in this study.  The grey line in each panel shows the best-fit Schechter function to the binned LF points using the $\chi^2$ grid approach.  The black line corresponds to the best-fit estimated from the MLE method.  The yellow shaded region represents $1\sigma$ uncertainties of the LF estimated from the MLE method.  The vertical black dotted lines denote the absolute $H$-band magnitude limit used for fitting the LF.  The best fitting $M_{H}^{*}$ of both methods and their $1\sigma$ uncertainties are denoted by the vertical arrows and the horizontal error bars.}  % while the pink shaded region represents the final $1\sigma$ uncertainties after taking into account the uncertainties due to field variation in the statistical background subtraction process.}
  \label{fig_lf_montage}
\end{figure*}

Figure~\ref{fig_lf_montage} shows the red sequence LF of the seven GOGREEN clusters.  The binned LFs are plotted to the respective cluster magnitude limits.  In general, the measured binned LFs can be described reasonably well by a single Schechter function.  For SpARCS1616, the apparent excess of galaxies that are brighter than the BCG is caused by the bright galaxies with comparable brightness with the BCG (see Figure~\ref{fig_cmd}) and the choice of binning.  On the other hand, the excess in SpARCS1634 is a result of an off-center galaxy that is brighter than the assigned BCG (see Section~\ref{subsec:Red sequence selection} for details).

In all the clusters there is a gradual decrease of the number of red sequence galaxies towards the faint end, which is also reflected in the derived $\alpha$s: all seven clusters show $\alpha \gtrsim -0.8$.  We find that the binning approach and the MLE approach give consistent estimates of Schechter parameters.  In all clusters the derived $\alpha$ and $M_{H}^{*}$ from both methods are consistent within $1\sigma$.  The best-fitting Schechter parameters of the MLE method ($\alpha$, $M_{H}^{*}$) and the effective number of red sequence galaxies ($N_{\rm{eff,RS}}$) that goes into the LF derivation for each cluster are given in Table~\ref{tab_fit_summary}. Note that due to statistical background subtraction, the effective number of red sequence galaxies in each cluster is no longer an integer.
%For clusters with low S/N, we find that the output of the binning approach depends heavily on the choice of binning. % (see Appendix X for a detailed comparison of the two approaches).
 %Previous works have also found a similar decline at the faint end of LF in clusters at lower redshifts \citep[e.g.][]{Rudnicketal2009, Martinetetal2015}.

%==== Individual LF fig =========                                      %%CHECKED_REFV2                                    
%==========================

%%%%%% Composite LF %%%%%%% 
\subsection{Composite luminosity function}                                   %CHECKED_REFV2  
\begin{figure}
  \centering
  \includegraphics[scale=0.517]{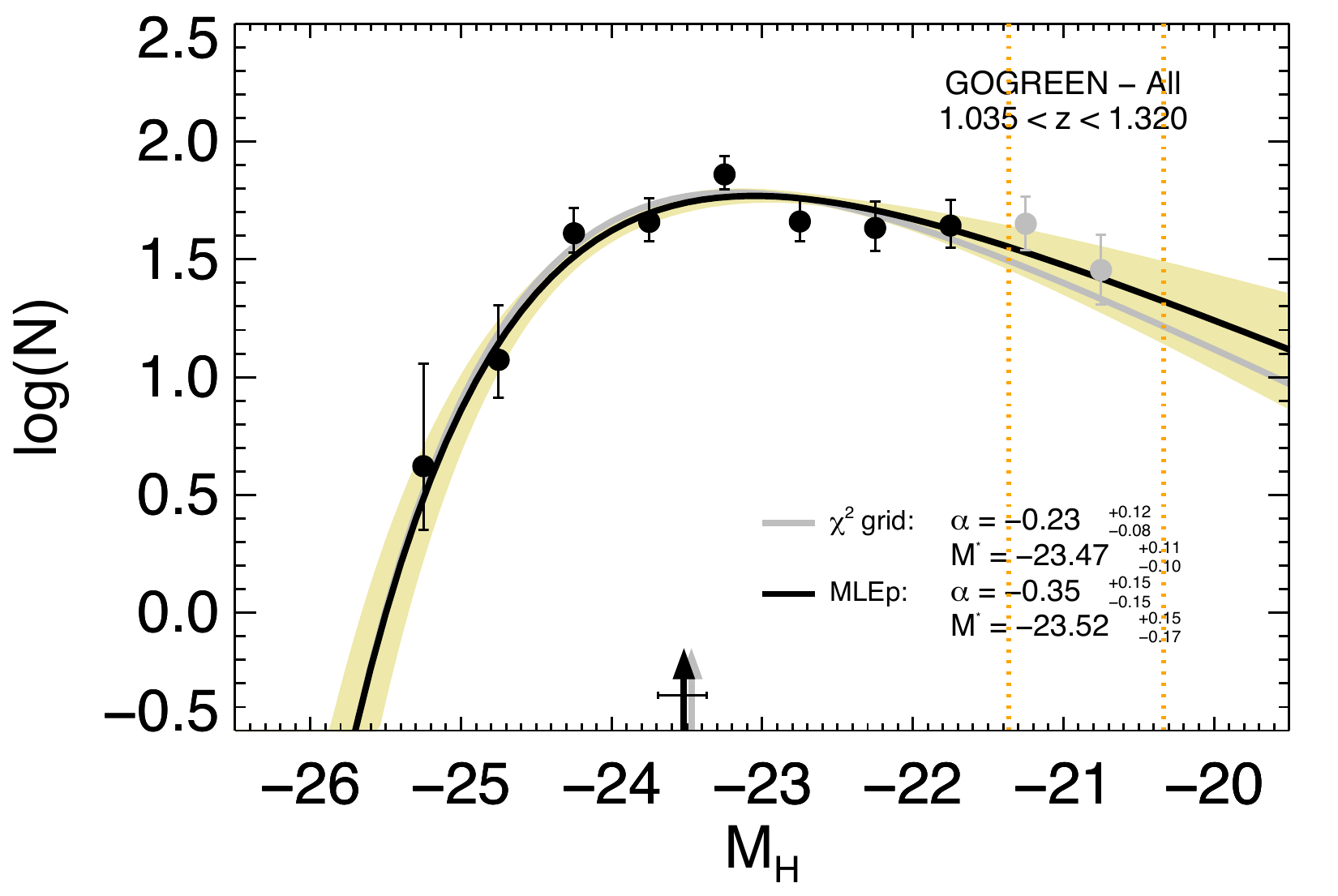}
  \caption{Composite rest-frame $H$-band red sequence LF of the seven GOGREEN clusters.  Passive evolution correction have been applied before stacking to bring all the clusters to the mean redshift of the sample ($\bar{z} \sim1.15$).  The orange dotted lines bracket the range of the absolute $M_H$ magnitude limits (after corrected to $\bar{z} \sim 1.15$) of the seven clusters.  The data points show the stacked LF using the \citet{Colless1989} method.  Points that are within the magnitude limits of all the clusters are shown in black and those that are within the range of magnitude limits are shown in grey.  The grey line shows the best-fit to the LF (black points) using the $\chi^2$ grid approach.  The black line corresponds to the best-fit estimated from the MLE method.  The yellow shaded region represents $1\sigma$ uncertainties of the LF estimated from the MLE method. Vertical arrows and horizontal error bars of the same color show the corresponding best-fit $M_{H}^{*}$ and $1\sigma$ uncertainty.}
  \label{fig_lf_stack_all}
\end{figure}

In Figure~\ref{fig_lf_stack_all} we show the composite red sequence LF of the seven GOGREEN clusters. The LFs of individual clusters are corrected to the mean redshift of the sample at $\bar{z}\sim1.15$ before stacking.  Note that the BCGs have been removed before deriving the LF.

The bright end of the LF appears to be well described by the exponential part of the Schechter function.  Previous studies at lower redshifts have reported an excess of red sequence galaxies at the bright end that deviates from the best fitting Schechter function \citep[e.g.][]{Bivianoetal1995, Barrenaetal2012, Martinetetal2015}.  Although part of the excess seen in previous works is due to the fact that these works included the cluster BCGs in the LF,  the excess has been shown to be made up by bright red sequence galaxies that are not BCGs \citep[see e.g.][]{Barrenaetal2012, Ceruloetal2016}.  We do not find evidence of such excess in the composite GOGREEN LF, although the bright end of our composite LF has large uncertainty due to the small number of bright galaxies we have in the sample (and small number of clusters) and the variation in the number of bright galaxies among the clusters.
%The large uncertainty at the bright end of our LF is due to the small number of bright galaxies we have in the sample (and small number of clusters) and the variation in the number of bright galaxies among the clusters.

By combining the sample as a whole, we can constrain the cluster-average $\alpha$ and $M_{H}^{*}$ simultaneously with higher accuracies.  The measured composite LF shows a prominent decline at the faint end, with best fitting Schechter parameters ${\alpha \sim -0.35^{+0.15}_{-0.15}}$ and ${M_{H}^{*} \sim -23.52^{+0.15}_{-0.17}}$ from the MLE method and ${\alpha \sim -0.23^{+0.12}_{-0.08}}$ and ${M_{H}^{*} \sim -23.47^{+0.11}_{-0.10}}$ from the $\chi^2$ grid method.\footnote{Note that both fits are fitted down to the brightest magnitude of the magnitude limits of the individual clusters. If the extra two data points that are within the range of magnitude limits are included in the $\chi^2$ grid fit (as permitted by the \citet{Colless1989} method for stacking LF with different limits), we find that the resultant ${\alpha \sim -0.42^{+0.06}_{-0.06}}$ and ${M_{H}^{*} \sim -23.63^{+0.08}_{-0.10}}$ differ by $> 1 \sigma$ with the above $\chi^2$ grid fitting results, but are consistent within $1\sigma$ of the MLE method.}  In the section below when comparing $\alpha$ and $M_{H}^{*}$ among samples we will mainly refer to those derived from the MLE method, as they give more conservative results.

We also study the halo mass dependence and cluster-centric radial dependence of the composite LF.  The results are shown in Appendix~\ref{app:Does the LF Depend on Halo mass} and~\ref{app:Is there any radial dependence}. We see no obvious dependence of Schechter parameters on halo mass, but there may be a hint of a radial dependence, in a way that the LF in the inner $0.5$ Mpc show a more positive $\alpha$ than the outer region $0.5 < R \leq 1.0$ Mpc. 

We have also investigated the potential effect of source blending in our photometry and verified that source blending is unlikely to affect our conclusions. The tests and results are described in Appendix~\ref{app:Potential effect of source blending on the LF}.

%==== Composite LF - all =====                                                      %CHECKED_REFV2 
%=========================

%%%%%% Comparison with low redshift sample %%%%%%%
\subsection{Comparison with low redshift sample}                              %CHECKED_REFV2
\label{subsec:Comparison with low redshift sample}
In this section we examine the redshift evolution of the cluster red sequence by comparing our results to the EDisCS sample at $\bar{z} = 0.60$.  Before comparing their LFs, we first check if the two samples are comparable in mass.  Various studies have shown that there are possible correlations between the LF parameters and cluster mass \citep[e.g.][]{Popessoetal2006, Martinetetal2015}, it is therefore important to also take into account the growth of the cluster when comparing clusters at different redshifts.

\begin{figure}
  \centering
  \includegraphics[scale=0.51]{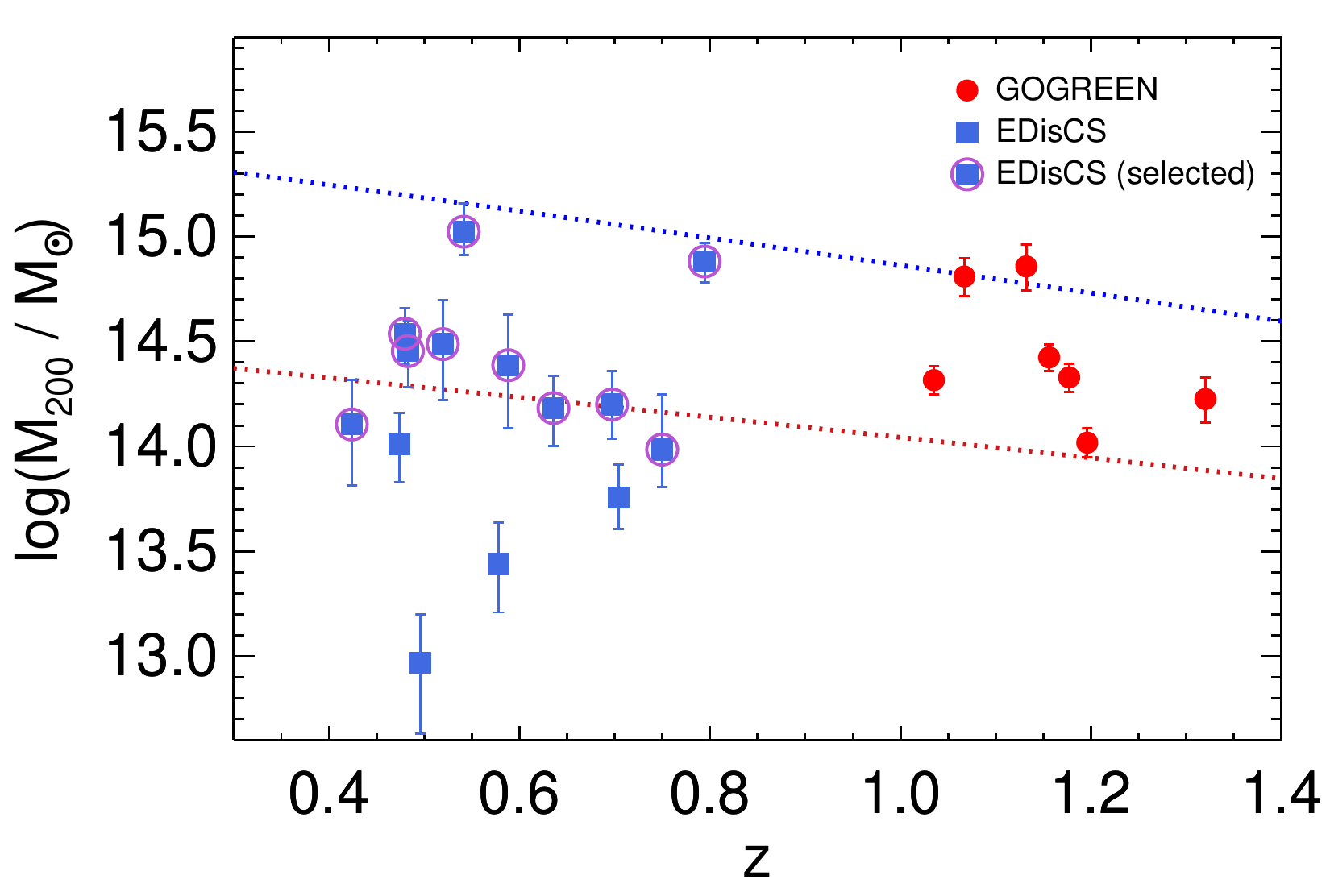}
  \caption{Halo masses of the GOGREEN and EDisCS clusters.  The blue and red dotted lines show the expected halo mass history of the most massive EDisCS cluster CL1232-1250 and the least massive GOGREEN cluster SpARCS1638, computed using the concentration-mass relation and mass accretion history code \citep[\textsc{commah},][]{Correaetal2015a,Correaetal2015b,Correaetal2015c}.  It can be seen that given the expected halo mass histories, ten of the EDisCS clusters (circled) are plausible descendants of the GOGREEN clusters.}
  \label{fig_m200_z}
\end{figure}

Figure~\ref{fig_m200_z} shows the halo mass of the clusters in the two samples with redshift.  We also plot the expected halo mass accretion history of the most massive EDisCS cluster (CL1232-1250) and the least massive GOGREEN cluster (SpARCS1638), computed using the concentration-mass relation and mass accretion history code \citep[\textsc{commah},][]{Correaetal2015a,Correaetal2015b,Correaetal2015c}.  \textsc{commah} uses the extended Press-Schechter formalism \citep[e.g.][]{Bondetal1991, LaceyCole1993} to compute the average halo mass accretion history for a halo with a given initial mass at a certain redshift.  Note that here we merely use the expected halo mass histories of the two clusters as a reference.  Due to the stochastic nature of structure formation, there is considerable scatter in the average mass accretion history \citep[$\sim0.2$ dex at $z\sim1$ as seen in simulations e.g.][]{vandenBoschetal2014} that is not included in the analytical models.  

Given the expected growth, it can be seen that while most of the EDisCS sample are plausible descendants of the GOGREEN clusters within the uncertainties, four of the clusters may have halo masses that are too low to compare with the GOGREEN sample. Therefore, in addition to the comparison with all fourteen EDisCS clusters, we also compose a subsample that comprises only the plausible descendants of the GOGREEN clusters (marked with circles in Figure~\ref{fig_m200_z}), for which we refer to as the selected EDisCS clusters below.  Overall, we find that comparing the GOGREEN clusters to all EDisCS clusters or the subsample that is restricted to plausible descendants results in consistent conclusions. For completeness, the results of both comparisons are shown and discussed below.

 %\footnote{Strictly speaking, the least massive EDisCS cluster (CL1420-1236) is below the expected halo mass history line of SPT0205.  Nevertheless, given the possible scatter we keep this cluster in the comparison. We have also verified that removing this cluster does not change the conclusion.}
%\footnote{The $M_{200}$ of SPT0205 from our dynamical analysis is preliminary and has large uncertainty. We notice that it is below $10^{14} M_{\odot}$ and is also lower than the mass found in \citet{Rueletal2014}.  However, as we show in Figure~\ref{fig_m200_z}, it will grow into a low-mass cluster that is comparable to our comparison sample in intermediate redshift.}
%EDISCS mean mass   2.6641929e+14   \pm 4.5390913e+13
%SpARCS mean mass   3.3537286e+14  \pm 8.8127512e+13  v2 SPT0205 mass
%SpARCS mean mass   3.3230e+14 \pm 3.3696e+13  v3 cluster masses

%==== M200 vs z for cluster selection ========                               %CHECKED_REFV2
%=====================================

\begin{figure*}
  \centering
  \includegraphics[scale=1.25]{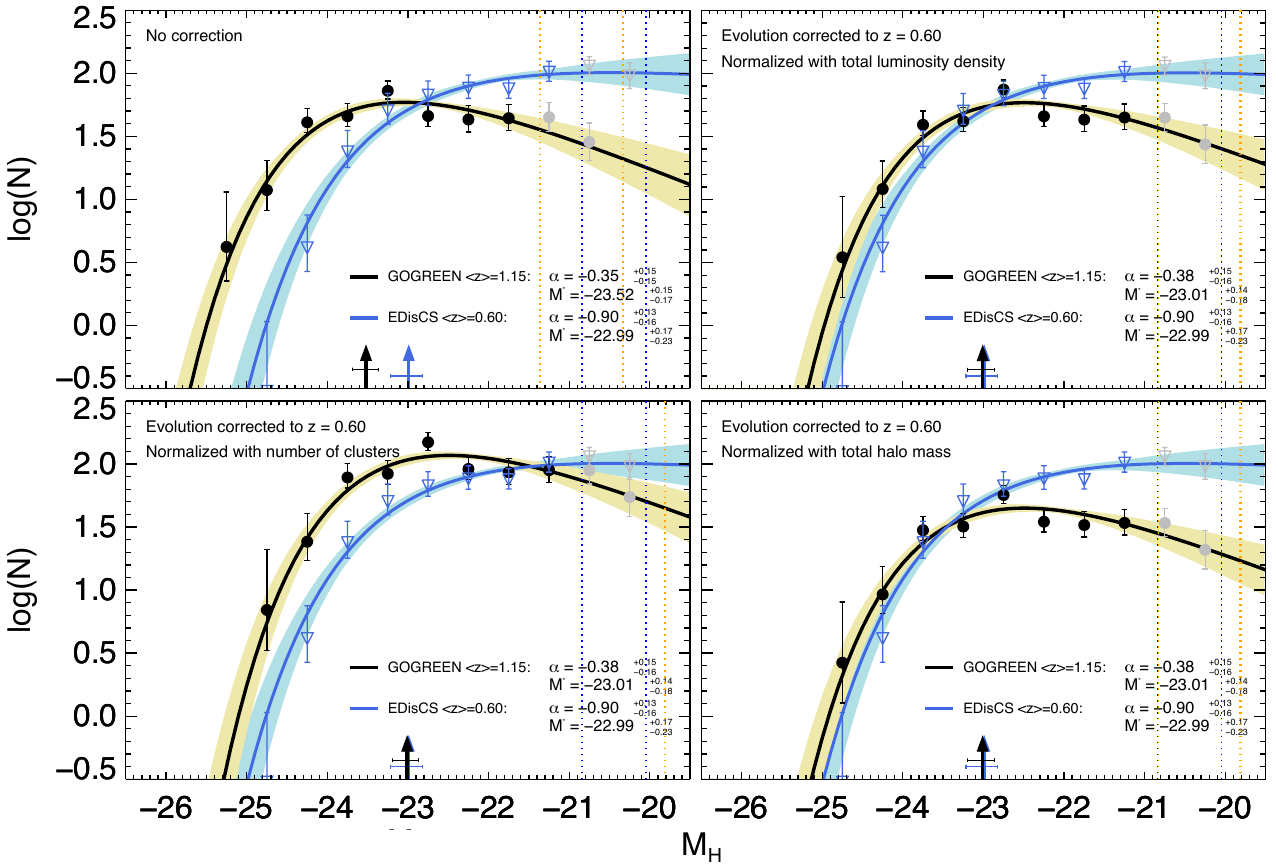}
  \caption{Composite rest-frame $H$-band red sequence LF of GOGREEN and EDisCS clusters.  Top left: composite LFs of the two samples at their respective mean redshifts without evolution correction and correction in the normalisation.  Top Right:  Passive evolution correction has been applied to the GOGREEN clusters to bring them to the mean redshift of the EDisCS sample.  The GOGREEN LF and its best-fits are scaled (i.e. shifted vertically) to have the same total luminosity density as the EDisCS LF.  Bottom left: same as top right, but the GOGREEN LFs are scaled to have the same number of clusters as the EDisCS sample.  Bottom right:  same as top right, but the GOGREEN LFs are rescaled to have the same total halo mass at $z\sim0.6$ as the EDisCS sample. The black and blue lines in all panels correspond to the best-fit Schechter function for the GOGREEN and the EDisCS sample, respectively.  Vertical arrows of the same color show the corresponding best-fit $M_{H}^{*}$. The yellow and cyan shaded region represents $1\sigma$ fitting uncertainties of the GOGREEN and EDisCS LF.  The orange and blue vertical dotted lines bracket the range of $M_H$ magnitude limits of the GOGREEN and EDisCS clusters. Due to the difference in mean $M_{200}$ mass of the two samples, we suggest that the LF should be renormalized with halo mass for comparison (see Section~\ref{subsec:Comparison with low redshift sample} for details). It can be seen that there is a lack of red sequence galaxies in the faint end of the GOGREEN LF compared to the EDisCS sample.}
  \label{fig_lf_zevo}
\end{figure*}

\begin{figure*}
  \centering
  \includegraphics[scale=1.25]{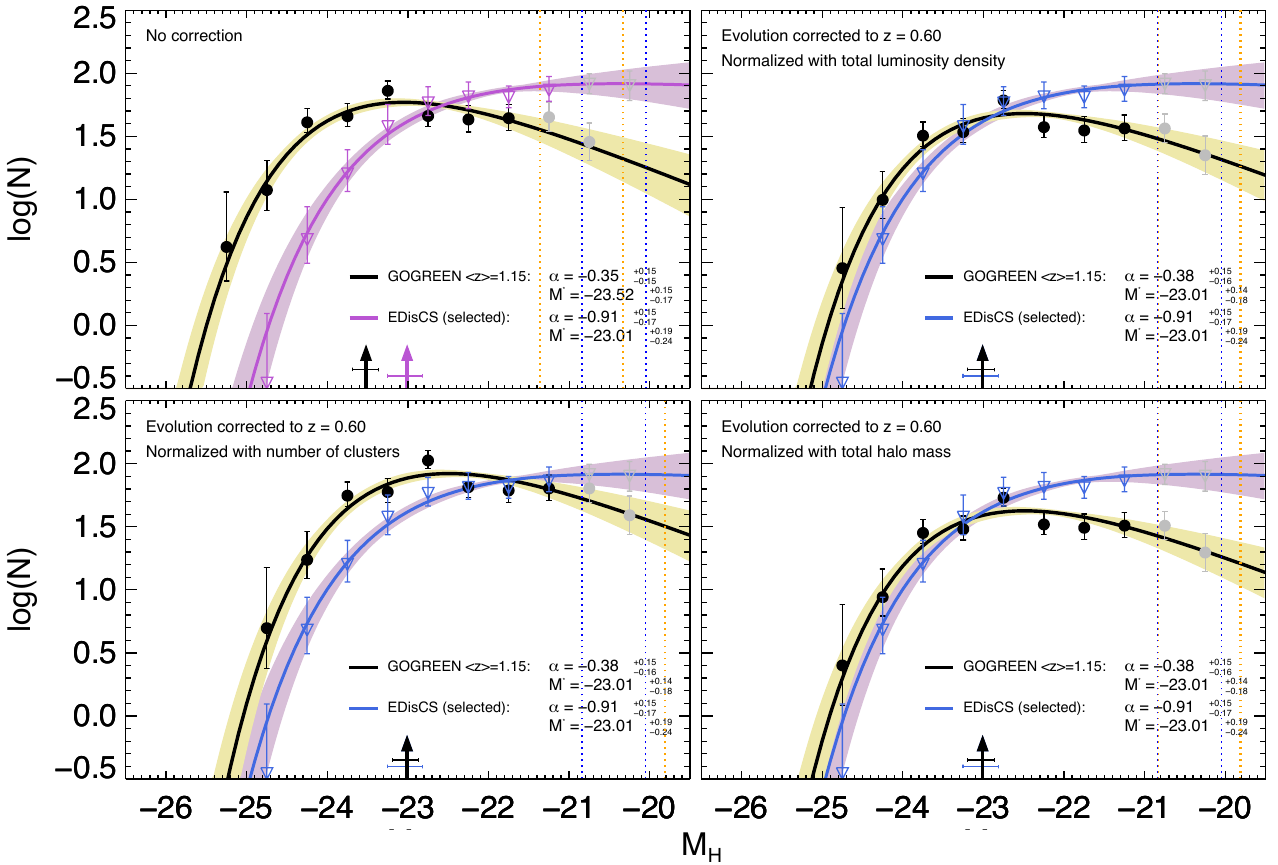}
  \caption{Composite rest-frame $H$-band red sequence LF of GOGREEN and the selected EDisCS clusters.  Same as Fig~\ref{fig_lf_zevo}, but the composite LF includes only ten clusters in the EDisCS sample that are plausible descendants of the GOGREEN clusters within the uncertainties (violet).}
  \label{fig_lf_zevo_esel}
\end{figure*}

The comparison of the composite LF of the GOGREEN clusters to all the EDisCS clusters and to the selected EDisCS clusters is shown in Figure~\ref{fig_lf_zevo} and Figure~\ref{fig_lf_zevo_esel}, respectively.  We have corrected the GOGREEN LFs for passive evolution to $z = 0.60$ to account for the fading of the stellar population.  The correction is again computed using BC03 SSP, assuming a formation redshift of $z_f = 3$ and solar metallicity.  In order to trace the redshift evolution of the LF,  a correction to the normalisation of the GOGREEN LF is also needed so that the (relative) number counts of the two samples can be compared directly.  In Figure~\ref{fig_lf_zevo} and~\ref{fig_lf_zevo_esel} we have rescaled the evolution corrected GOGREEN LF in three different ways, such that it has the same total luminosity density, number of clusters and total halo mass at $z=0.6$ as the EDisCS LFs, respectively.  In previous works the LFs being compared are often rescaled to have total luminosity density.  This is, however, only useful in comparing the shape (i.e. $\alpha$ and $M_{H}^{*}$) of the LFs as it provides no information on the evolution in absolute galaxy numbers.  One way is to rescale with the number of clusters as shown in the bottom left panel, but the results can be biased by the mass distribution of the samples even if the samples have been shown to be plausibly evolutionarily linked. 

Despite the fact that we have twice as many clusters in the EDisCS sample as in GOGREEN, the mean $M_{200}$ of the EDisCS sample ($2.7 \pm 0.5 \times 10^{14} M_{\odot}$) is lower than the mean $M_{200}$ of GOGREEN (${3.3 \pm 0.3 \times 10^{14} M_{\odot}}$) before even taking into account the growth of the cluster mass over the redshift range. Similarly, the mean $M_{200}$ ($3.5 \pm 0.7 \times 10^{14} M_{\odot}$) of the selected EDisCS clusters is already comparable to the GOGREEN sample.\footnote{We note that the mean $M_{200}$s of the full EDisCS sample and the selected EDisCS cluster subsample are comparable within $1\sigma$ with GOGREEN. Nevertheless, this will not be true at $z=0.6$ as GOGREEN clusters will grow by a factor $\sim2.1$ according to the expected halo mass accretion history from \textsc{commah}.}  It is therefore potentially problematic to rescale the LF using solely the ratio of the number of clusters in the EDisCS and GOGREEN samples in our case, as the difference in the mass distributions in the samples might lead to incorrect conclusions.  To solve this we rescale the GOGREEN LF using the ratio of the total halo mass of the two samples at $z=0.6$ (bottom right panel in both figures).  The expected average halo mass of the GOGREEN sample at $z=0.6$ ($7.0 \times 10^{14} M_{\odot}$) is again estimated with \textsc{commah}.  This is essentially halo-mass matching the samples and comparing their LFs per unit halo mass.  Comparing with the result that rescales with number of clusters, it can be seen that rescaling with halo-mass gives a smaller normalisation, which is due to the fact that GOGREEN clusters are on average more massive than EDisCS clusters.

From Figure~\ref{fig_lf_zevo}, it is clear that the bright ends of the GOGREEN and EDisCS LFs are consistent with each other after we correct for the passive evolution of the GOGREEN clusters to the EDisCS redshift and halo-mass match the two samples.  This suggests that the bright red sequence population in the cluster is mostly in place already at $z\sim1$.  We can also see from Figure~\ref{fig_lf_zevo_esel} that restricting the sample to only plausible descendants gives consistent results.

\begin{figure}
  \centering
  \includegraphics[scale=0.535]{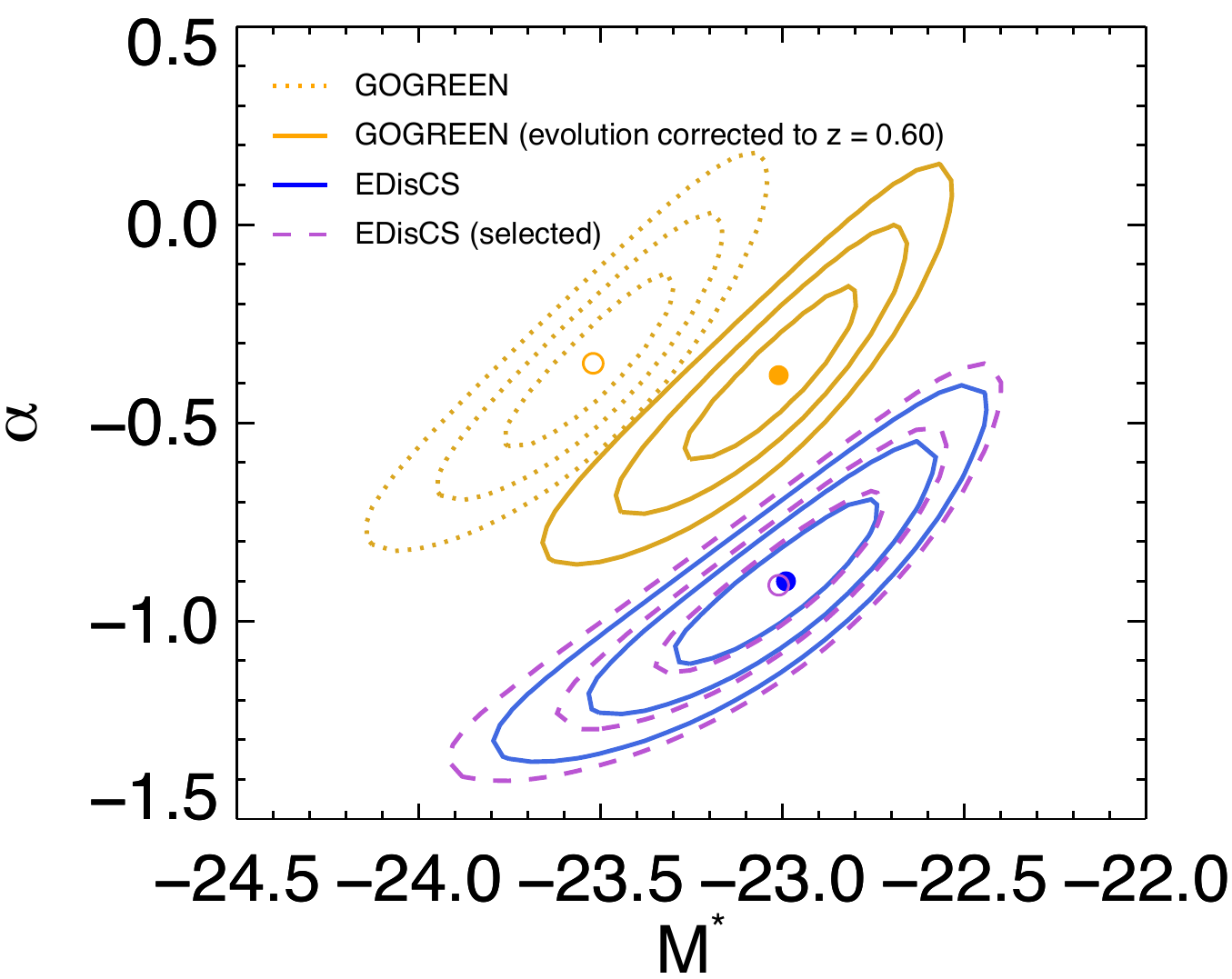}
  \caption{The likelihood contours for the Schechter parameters $\alpha$ and $M_{H}^{*}$ for the GOGREEN and EDisCS clusters from the MLE approach. Dotted and solid orange lines show the 1,2 and 3$\sigma$ confidence contours of the two parameters before and after applying evolution correction for the GOGREEN sample.  Solid blue and dashed violet lines show the confidence contours of the full EDisCS sample and selected EDisCS cluster sample.  The circles correspond to the best fitting Schechter parameters. There is a $>3\sigma$ level difference between the Schechter parameters of the evolution corrected GOGREEN sample  and both EDisCS samples, which primarily comes from $\alpha$.}
  \label{fig_lf_zevo_contour}
\end{figure}

The faint end, on the other hand, shows an evolution from $z\sim1.15$ to $z\sim0.60$.  This is evident from the lack of red sequence galaxies fainter than $M_{H}^{*}$ in the GOGREEN LF compared to the EDisCS LF.  Since the two Schechter parameters $\alpha$ and $M_{H}^{*}$ are degenerate, in Figure~\ref{fig_lf_zevo_contour} we show the likelihood contours for both parameters from the MLE approach. It can be seen that there is a $>3\sigma$ level difference between the Schechter parameters of the evolution corrected GOGREEN sample and both EDisCS samples. Marginalising the parameters suggests the difference primarily comes from $\alpha$, while the evolution corrected $M_{H}^{*}$ of the two LFs are well consistent within $1\sigma$ and show no evolution.  

Our result of an evolving $\alpha$ with redshift is consistent with recent studies by \citet{Zhangetal2017} and \citet{Sarronetal2018} at lower redshift ranges. \citet{Zhangetal2017} studied the red sequence LF for a sample of X-ray selected clusters from the Dark Energy Survey at $0.1 < z < 1.05$ and reported a $\sim1.9\sigma$ redshift evolution in the faint end slope. \citet{Sarronetal2018} studied the evolution of the LF of both early-type and late-type galaxies in clusters at $z \leq 0.7$ and found an increase in faint galaxies in both populations with decreasing redshift. Our result is also consistent with the `downsizing' scenario \citep{Cowieetal1996} of red sequence formation seen in low redshift clusters \citep[e.g.][]{Nelanetal2005, Smith2005, Smithetal2012}, such that faint red-sequence galaxies become quiescent at a later time than the bright massive ones.

Besides LFs that are computed within a physical radius of $0.75$Mpc, we also compare GOGREEN and EDisCS LFs that are computed within $0.5 R_{200}$.  We find the same conclusion as found using $R \leq 0.75$Mpc, that the bright ends of the two $0.5 R_{200}$ LFs are consistent with each other, and that the faint end slope shows an evolution from $z\sim1.15$ to $z\sim0.60$. The best-fitting Schechter parameters are given in Table~\ref{tab_fit_summary}.

%==== Composite LF - redshift evolution =====                                   %CHECKED_REFV2
%====================================

%==== Composite LF - redshift evolution (EDisCS selected) =====        %CHECKED_REFV2
%====================================

%==== Alpha - M* for comparison============                                 %CHECKED_REFV2
%====================================

%%%%%% The total red sequence luminosity %%%%%%%    
\subsection{The total red sequence luminosity}                                      %CHECKED_REFV2
\label{subsec:The total red sequence luminosity}
To quantify the build up of the faint end of the red sequence, in this section we measure the growth of the total luminosity of the red sequence over time.\footnote{Although we do not convert the total red sequence luminosity to a total mass in this section, we would like to inform the reader that the rest-frame $M/L_H$ at $z=0.60$ is $\sim0.63$, as derived from SSP model assuming $z_f = 3.0$ and $Z=Z_\odot$.}  We integrate the best fitting LFs of the GOGREEN and the full and selected EDisCS samples, as well as the individual cluster LFs.  To make sure the conclusion is not affected by the extrapolation of the LF, we have also integrated the LFs down to the magnitude limit of the EDisCS sample ($M_H \leq 21$).  The results are consistent with the complete integrals.

\begin{figure}
  \centering
  \includegraphics[scale=1.000]{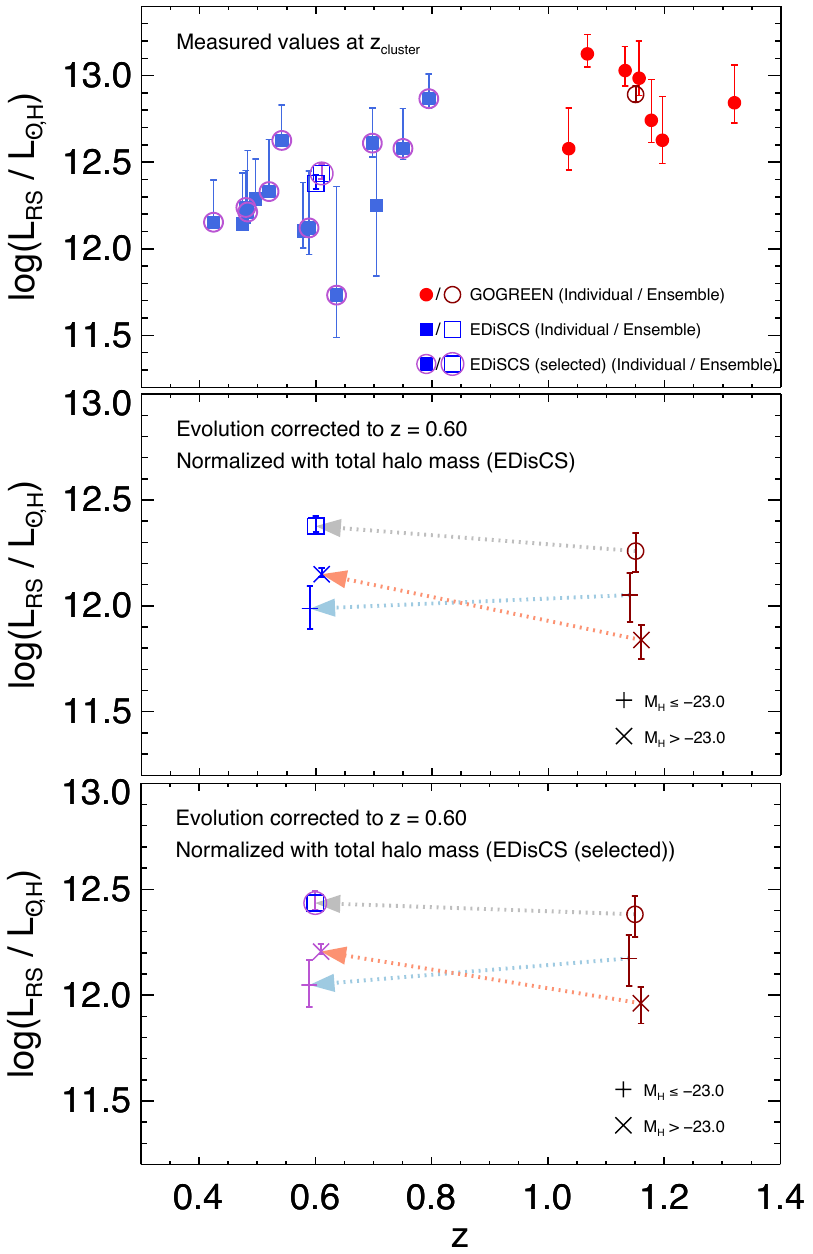}
  \caption{Total red sequence luminosity of the GOGREEN and EDisCS clusters. Top: luminosities measured by integrating the LF (without evolution correction) at the respective cluster redshift.  Middle: Passive evolution correction has been applied to fade the mean red sequence luminosity of the GOGREEN sample to the mean redshift of the full EDisCS sample, and the mean luminosities of the GOGREEN sample are rescaled to have the same total halo mass at $z\sim0.6$ as EDisCS.  The mean red sequence luminosity of the two samples are derived from the composite LFs. The pluses and crosses correspond to the red sequence luminosity of each sample with $M_H \leq -23.0$ (bright end) and $M_H > -23.0$ (faint end), respectively.  Bottom:  Same as the middle panel, except that the selected EDisCS sample is used for rescaling. It can be seen that the majority of the growth in luminosity occurs at the faint end of the red sequence.}
  \label{fig_lf_rslumin_z}
\end{figure}

Figure~\ref{fig_lf_rslumin_z} shows the total red sequence luminosity $L_{\rm{RS}}$ of the samples.  Without any corrections, the measured values of the red sequence luminosity of the GOGREEN clusters are higher than both the full and selected EDisCS cluster samples (top panel).  After accounting for the fading of the stellar population with the passive evolution correction and rescaling the GOGREEN LF using the ratio of the total halo mass of the full and selected EDisCS samples (as in bottom right panel of Figure~\ref{fig_lf_zevo} and ~\ref{fig_lf_zevo_esel}), in the middle and bottom panels we see an evolution in the (mean) red sequence luminosities between the GOGREEN sample to the EDisCS samples. Taking into account the uncertainties, the red sequence luminosities only grow by $\sim31\% \pm 30\%$ (full EDisCS sample, $\sim13\% \pm 28\%$ if we consider the selected EDisCS sample) over the $\sim2.6$ Gyrs between $z \sim 1.15$ and $z \sim0.60$.

It is not surprising that the growth of red sequence luminosities between the two samples is not significant, as we have shown that the bright end of the LF, which constitutes the majority of the red sequence light, is fully assembled in the GOGREEN sample.  If we split the red sequence luminosities at $M_H = -23$ (passive evolution corrected), which roughly corresponds to $M_{H}^{*}$ of both samples at $z\sim0.60$, we find that the mean luminosity of the bright end ($M_H \leq -23$) of the GOGREEN clusters is consistent with EDisCS ($\sim -13\% \pm 31\%$ (full), $\sim -24\% \pm 29\%$ (selected), pluses in Figure~\ref{fig_lf_rslumin_z}), albeit with considerable uncertainty.  On the other hand, the faint end ($M_H > -23$) of the GOGREEN sample needs to grow by almost a factor of $2$ ($\sim104\% \pm 39\%$ (full), $\sim76\% \pm 35\%$ (selected)) in luminosity to meet the EDisCS sample (crosses).

%==== RS lumin - redshift evolution =========                                    %CHECKED_REFV2
%====================================

Our result that the bright end is already assembled by $z\sim1.15$ is consistent with \citet{vanderBurgetal2015}, in which they compared the stellar mass density distributions of $z\sim1$ GCLASS clusters, four of which are also in the sample used in this work, to a sample of low redshift clusters ($0.04 < z < 0.26$).  They found that the central parts ($R < 400$ kpc) of the stellar mass distributions of satellite galaxies in local galaxy clusters are already in place at $z\sim1$.  Here we show that this is also true for red sequence galaxies as the total red sequence luminosity is mostly in place from the contribution of the assembled bright end.  Interestingly, there has been evidence showing that even the bright end is still being assembled at higher redshift.  For example, \citet{Rudnicketal2012} derived the red sequence LF of a $z=1.62$ cluster IRC0218 and found that the bright end of the LF is less populated than its descendants at $z\sim 0.6$ with a factor of two less in the total red sequence luminosity despite the fact that the shape of the LFs are consistent with each other.  If IRC0218 is a typical case, this is then consistent with the picture that the massive end evolves faster than the faint end.  The massive end completes most of the assembly until $z\sim1$ and then followed up by the build up of the faint end of the red sequence.  Note that this does not mean that the bright end of the red sequence completely stops evolving after $z\sim1$, as other processes, such as the growth of the BCGs via mergers and the arrival of newly quenched galaxies, will continue to affect the bright end of the red sequence to the present day (see Section~\ref{subsec:Growth of the BCG and ICL} for an estimation of the effect of BCG growth to the total red sequence luminosity).

The comparison here, however, assumes that all red sequence galaxies evolve passively in the same way down to $z \sim 0.60$ and the average growth of the GOGREEN sample is well described by the average mass accretion history, while in reality cluster galaxies and the clusters themselves may have different formation histories.  We check that varying $z_f$ in the SSP models introduces $\sim0.1$ mag variation in the evolution correction. This corresponds to a $\sim 0.05$ dex (i.e. $\sim11-12\%$) systematic uncertainty in the derived mean red sequence luminosities of the GOGREEN clusters and is $\sim3-4$ times smaller than our uncertainties.  Note that the LF and the total red sequence luminosity we show here are computed within a physical radius of $0.75$Mpc.  We also verify that using LFs that are computed within $R \leq 0.5 R_{200}$ gives a similar conclusion, in the sense that the majority of the growth in the red sequence luminosity occurs at the faint end.

%===== Alpha M* Table===================                             %CHECKED_REFV2
\begin{table*}
  \caption{Effective number of red sequence galaxies and the derived Schechter parameters of the red sequence LFs with the MLE method.}
  \centering
  \label{tab_fit_summary}
  \begin{tabular}{lcccc}
%  \begin{tabular}{@{}lcccccc@{}}  
  \hline
  \hline
%Name &  Redshift & $\alpha$\textsuperscript{a} & $M_{H}^{*}$\textsuperscript{b}      \\
Name &  Redshift &   $N_{\rm{eff,RS}}$\textsuperscript{a} &  $\alpha$ & $M_{H}^{*}$    \\
             &     $z$     &                                            &   (mag)                            \\
  \hline
  SpARCS1051  & $1.035$ &  ${37.0 \pm 1.3}$    &   ${-0.41^{+0.32}_{-0.37}}$  &     ${-23.17^{+0.36}_{-0.60}}$          \\  [0.5ex]    %UPDATED V528 REFV2
  SPT0546         & $1.067$ &   ${104.5 \pm 2.1}$  &  ${-0.41^{+0.20}_{-0.23}}$   &     ${-23.35^{+0.23}_{-0.32}}$         \\  [0.5ex]    %UPDATED V528 REFV2
  SPT2106         & $1.132$ &   ${71.6\pm 1.8}$   &  ${-0.37^{+0.25}_{-0.27}}$   &     ${-23.47^{+0.28}_{-0.38}}$          \\  [0.5ex]    %UPDATED V528 REFV2
  SpARCS1616  & $1.156$ &   ${57.0 \pm 1.4}$  &   ${-0.83^{+0.21}_{-0.28}}$   &     ${-24.16^{+0.35}_{-0.65}}$          \\  [0.5ex]    %UPDATED V528 REFV2
  SpARCS1634  & $1.177$ &   ${31.4 \pm 1.1}$  &  ${-0.42^{+0.35}_{-0.43}}$    &     ${-23.73^{+0.39}_{-0.69}}$          \\  [0.5ex]    %UPDATED V528 REFV2
  SpARCS1638  & $1.196$ &  ${29.5 \pm 1.7}$ &  ${-0.24^{+0.43}_{-0.50}}$   &     ${-23.33^{+0.40}_{-0.68}}$          \\  [0.5ex]    %UPDATED V528 REFV2
  SPT0205         & $1.320$ &   ${27.7 \pm 1.5}$  &  ${0.19^{+0.52}_{-0.58}}$    &     ${-23.47^{+0.36}_{-0.56}}$          \\   [0.5ex]    %UPDATED V528 REFV2
   \hline
%  All clusters ($R \leq 0.75$ Mpc) &  $1.035 \leq z \leq 1.320$   &     $-0.36^{+0.15}_{-0.15}$     &      $-23.51^{+0.15}_{-0.18}$          \\  [0.5ex]    %UPDATED V518
%  Low-mass clusters ($R \leq 0.75$ Mpc)  &  $1.035 \leq z \leq 1.320$       &     $-0.39^{+0.20}_{-0.26}$     &     $-23.62^{+0.21}_{-0.26}$            \\  [0.5ex]    %UPDATED V518
%  High-mass clusters ($R \leq 0.75$ Mpc)  &  $1.067 \leq z \leq 1.132$      &     $-0.46^{+0.13}_{-0.15}$     &     $-23.51^{+0.17}_{-0.23}$            \\  [0.5ex]    %UPDATED V518
  All clusters ($R \leq 0.75$ Mpc)               &  $1.035 \leq z \leq 1.320$      &   ${315.9 \pm 3.5}$  &   ${-0.35^{+0.15}_{-0.15}}$     &      ${-23.52^{+0.15}_{-0.17}}$           \\  [0.5ex]    %UPDATED V528 REFV2
  All clusters ($R \leq 0.5 R_{200}$)           &  $1.035 \leq z \leq 1.320$      &   ${250.8 \pm 2.3}$  &   ${-0.31^{+0.17}_{-0.18}}$     &     ${-23.42^{+0.16}_{-0.19}}$            \\  [0.5ex]    %UPDATED V535 REFV2
  High-mass clusters ($R \leq 0.75$ Mpc)  &  $1.067 \leq z \leq 1.132$      &   ${173.6 \pm 2.8}$  &  ${-0.33^{+0.17}_{-0.17}}$     &     ${-23.40^{+0.18}_{-0.23}}$            \\  [0.5ex]    %UPDATED V528 REFV2
  Low-mass clusters ($R \leq 0.75$ Mpc)   &  $1.035 \leq z \leq 1.320$      &   ${163.1 \pm 2.6}$  &   ${-0.43^{+0.19}_{-0.21}}$     &     ${-23.67^{+0.20}_{-0.26}}$            \\  [0.5ex]    %UPDATED V528 REFV2
  All clusters ($R \leq 0.5$ Mpc)                  &  $1.035 \leq z \leq 1.320$      &  ${214.9 \pm 3.5}$  &   ${-0.17^{+0.19}_{-0.20}}$     &      ${-23.38^{+0.17}_{-0.20}}$          \\  [0.5ex]     %UPDATED V533 REFV2
  All clusters ($R \leq 1.0$ Mpc)                  &  $1.035 \leq z \leq 1.320$      &  ${334.0 \pm 4.6}$  &   ${-0.32^{+0.15}_{-0.16}}$      &     ${-23.47^{+0.14}_{-0.17}}$          \\  [0.5ex]     %UPDATED V533 REFV2
  All clusters ($0.5 \leq R \leq 1.0$ Mpc)    &   $1.035 \leq z \leq 1.320$      &   ${119.2 \pm 2.9}$  &   ${-0.57^{+0.22}_{-0.26}}$      &     ${-23.64^{+0.25}_{-0.36}}$          \\  [0.5ex]     %UPDATED V523 REFV2
  EDisCS clusters  ($R \leq 0.75$ Mpc)    &  $0.424 \leq z \leq 0.794$      &   ${433.1 \pm 5.4}$   &  ${-0.90^{+0.13}_{-0.16}}$     &     ${-22.99^{+0.17}_{-0.23}}$          \\  [0.5ex]   %UPDATED V42 (All 14 clusters) REFV2
  EDisCS clusters (selected)  ($R \leq 0.75$ Mpc)    &  $0.424 \leq z \leq 0.794$      &  ${350.8 \pm 4.7}$ &     ${-0.91^{+0.15}_{-0.17}}$     &     ${-23.01^{+0.19}_{-0.24}}$          \\  [0.5ex]   %UPDATED V43 (Selected 10 clusters) REFV2
  EDisCS clusters  ($R \leq 0.5 R_{200}$)    &  $0.424 \leq z \leq 0.794$   &   ${328.3 \pm 4.4}$  &    ${-0.80^{+0.16}_{-0.18}}$     &     ${-22.85^{+0.19}_{-0.23}}$         \\  [0.5ex]   %UPDATED V46 (All 14 clusters) REFV2
  EDisCS clusters (selected) ($R \leq 0.5 R_{200}$)    &  $0.424 \leq z \leq 0.794$      &   ${280.1 \pm 4.2}$  & ${-0.82^{+0.17}_{-0.20}}$     &     ${-22.84^{+0.20}_{-0.27}}$         \\  [0.5ex]   %UPDATED V47 (Selected 10 clusters) REFV2
  \hline
    \multicolumn{5}{p{.7\textwidth}}{\textsuperscript{a} The effective number of red sequence galaxies ($N_{\textrm{eff,RS}}$) are computed by deriving 1000 realizations of the red sequence with the galaxy membership probabilities and calculating the sum of the probabilities in each realizations. $N_{\textrm{eff,RS}}$ is taken to be the median of the sums of the probabilities in these realizations and its uncertainty is the $1\sigma$ variation of the sums.} \\
\end{tabular}
\end{table*}
%======================================

%%%%%% Additional caveats %%%%%%%  
\subsection{Additional caveats}
\subsubsection{Growth of the BCG and ICL}                                        %CHECKED_REFV2
\label{subsec:Growth of the BCG and ICL}
There are two relevant components that we did not consider in the above comparisons, the growth of the BCG and the intracluster light (ICL). Recent studies on BCG stellar mass growth revealed that BCGs grow by a factor of $\sim2-3$ since $z\sim1$ primarily via major mergers, although when this growth takes place is still controversial \citep[e.g.][]{Lidmanetal2012, Linetal2013, Ascasoetal2014, Bellstedtetal2016}.  For the ICL, several simulation \citep[e.g.][]{Marteletal2012, Continietal2014} and observational works \citep[e.g.][]{DeMaioetal2015, DeMaioetal2018} have demonstrated that it originates from the disruption and tidal stripping of massive satellite galaxies of $\log(M / M_{\odot}) \sim 10-11$, and that the majority of the ICL growth happens below $z\sim1$.  Since the BCGs are excluded from our LFs, both the growth of the BCGs and ICL will manifest as a decrease in total red sequence luminosity at low redshift and therefore reduces the observed growth in total red sequence luminosity between the two samples.

A detailed analysis of the growth of the BCGs between the two samples is out of the scope of this paper, but we can roughly estimate the effect of BCG growth on the total red sequence luminosity using literature values.  Using the mean total (aperture) magnitudes of the GOGREEN BCGs\footnote{We are aware that the total magnitude is not an accurate measurement of the stellar component of the BCG but includes BCG+ICL, as we did not separate the ICL from the contribution of the outer halo of the BCG, although the BCG accounts for the bulk of the BCG+ICL mass at $z\sim1$ \citep{Continietal2018}. Here we simply use the luminosity of the BCGs to show that the effect of BCG growth to the red sequence luminosity is negligible.}, we find that on average the luminosities of the GOGREEN BCGs are $\sim10\%$ ($\sim3\%$) of the total red sequence luminosities.  Hence the growth of the BCGs would corresponds to $\sim3\%$ ($\sim1\%$) of the total red sequence luminosity, assuming they grow uniformly in time.  Similarly the contribution of the growth of ICL to the red sequence luminosity is also not significant; simulations show that most of the ICL assembles after $z\sim1$ and its fraction of mass grows from $\sim5-10\%$ to $\sim30-40\%$ of its $z\sim0$ value between $z\sim1.15$ and $z\sim0.60$ \citep{Continietal2014, Continietal2018}. They show that the stellar mass growth of the ICL has a BCG mass dependence.  The growth can reach $\sim1$ BCG mass by $z\sim0.5$ for more massive BCGs ($\log(M/M_{*}) > 11.5$, less growth for less massive BCGs).  Hence even if we assume all the ICL comes from red sequence galaxies, this still corresponds to $<10\%$ of the total red sequence luminosity.

\subsubsection{Contamination from dusty star forming galaxies}                               %CHECKED_REFV2
\label{subsec:Contamination from dusty star forming galaxies}
Another caveat comes from the contamination from dusty star forming galaxies.  Our red sequence selection alone is not able to differentiate between quiescent galaxies and star forming galaxies that appear red because of dust extinction.  In the last decade, the $UVJ$ color classification, which utilises the rest-frame $U-V$ and $V-J$ colors, has become a popular technique as it is able to separate `genuine' quiescent galaxies from dusty star-forming ones \citep[e.g.][]{Labbeetal2005, Wolfetal2005, Williamsetal2009, Muzzinetal2013b}.  By using the $UVJ$ classification we can estimate the average fraction of contamination in our red sequence, assuming the red sequence selection includes all the $UVJ$ quiescent and dusty star forming galaxies.  We utilise the existing multi-band catalogues of the GCLASS clusters and SpARCS clusters at a similar redshift range \citep[i.e. the $0.86 < z < 1.1$ and $1.1 < z <1.4$ bin of the cluster sample in][]{Nantaisetal2017}.  To select dusty star forming galaxies we follow the criterion in \citet{Spitleretal2014} to select objects in the $UVJ$ star forming region that have $V-J >1.2$.  We also tried the selection in \citet{Fumagallietal2014} (i.e. $U-V > 1.5$) and found that the two selection criteria gave consistent results for the clusters.

We find that dusty star forming galaxies contribute to $17\% \pm 6\%$ in number of the total galaxy population (for galaxies with $\log(M_{*}/M_{\odot}) \geq 10.3$) at $1.1 < z <1.4$.  For the $0.86 < z < 1.1$ bin the contribution is even lower, with $9\% \pm 9\%$.  This, combined with the fact that the quiescent fraction is $\sim 80\%$ at  $1.1 < z <1.4$ and $\sim85\%$ at $0.86 < z < 1.1$ \citep[][]{Nantaisetal2017}, suggests that the contamination from dusty star forming galaxies can contribute up to $\sim 11-20\%$ to our red sequence (at least for $M_H \geq -21.5$ at $z\sim0.65$, which the above mass limit roughly corresponds to).  It is likely that this contamination is even lower in the EDisCS sample, as the quiescent fraction in clusters increase with decreasing redshift and the fraction of dusty star forming galaxies decrease with decreasing redshift \citep[see also][for similar trends in the field]{Martisetal2016}.  Hence taking this into account will result in an even larger difference between the LF of the two samples than the one we see in Section~\ref{subsec:Comparison with low redshift sample} and \ref{subsec:The total red sequence luminosity}.

%We notice that fraction of dusty star forming galaxies is much higher in the field.  In fact, using the same selection on the COSMOS/UltraVISTA catalogue \citep{Muzzinetal2013c} we find that dusty star forming galaxies contribute to $32\% \pm 3\%$ of the total population at $1.1 < z <1.4$ ($26\% \pm 3\%$ at $0.86 < z< 1.1$).  %Therefore, using red sequence selection alone to compare the cluster and field population is potentially problematic.  %For this exact reason the comparison of the red sequence LF of the clusters to the field will be presented in a forthcoming paper.

%%%%%%%%%%%%%%%%%%%%%%%%%%%%%%%%%%%%%%%%%%%
%%%%%%%%%     Discussion      %%%%%%%%%%%%%%%%%%%%%%%%%
%%%%%%%%%%%%%%%%%%%%%%%%%%%%%%%%%%%%%%%%%%%
\section{Discussion}                                                                                            %CHECKED_REFV2
\label{sec:Discussion}
By comparing the GOGREEN sample with the EDisCS sample, we have shown that the red sequence cluster LF exhibits an evolution in the faint end slope $\alpha$ from $z\sim1.15$ to $z\sim0.60$.  The bright end, on the other hand, seems to be mostly in place already by $z \sim1.15$, which suggests a strong luminosity (or mass) dependence in the build up of the red sequence.  With only two samples, however, we are not able to truly trace the redshift evolution of the build up of the faint end.  In the following sections we investigate the redshift dependence of the build up of the red sequence and how this build up depends on the environment.

%%%%%% The evolution of the cluster faint-to-luminous ratio %%%%%%%  
\subsection{The evolution of the cluster faint-to-luminous ratio}                         %CHECKED_REFV2
\begin{figure*}
%  \centering
  \includegraphics[scale=1.125]{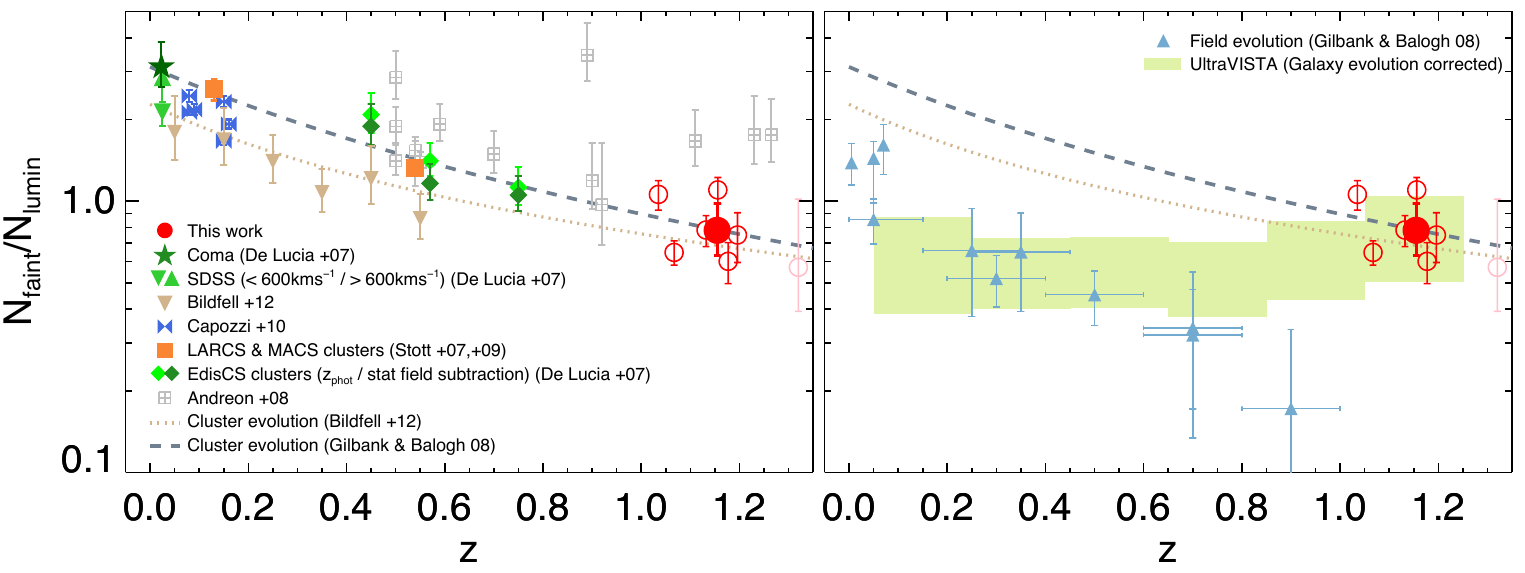}
  \caption[DGR]{The red sequence faint-to-luminous ratio ($N_{\rm{faint}}/N_{\rm{lumin}}$) of the GOGREEN clusters.  Left: Comparison with the literature values of cluster sample at different redshifts.  Empty symbols correspond to the ratios of individual clusters \citep[GOGREEN \&][]{Andreonetal2008}, while filled symbols correspond to the ratios deriving from cluster samples \citep[][]{DeLuciaetal2007, Stottetal2007, Stottetal2009, Capozzietal2010, Bildfelletal2012}.  Due to insufficient depth of the image, the ratio of SPT0205 is computed using extrapolation of the fitted LF and is shown in pink (see Section~\ref{subsec:Faint-to-luminous Ratio} for details). The red filled circle shows the ratio derived using composite LF of all seven clusters.  It is evident that there is a general trend of decreasing faint-to-luminous ratio with increasing redshift up to $z\sim1.15$. The ratio of the GOGREEN clusters are consistent with the evolution relations derived by \citet{GilbankBalogh2008} and \citet{Bildfelletal2012}, respectively.  Right: Comparison with the field.  The light green shaded regions correspond to the $1\sigma$ regions of the field faint-to-luminous ratios derived with the UltraVISTA catalogue, taking into account the variation in the evolution correction from different assumptions (see Section~\ref{sec:Comparison to the field} for details).  The field ratios in \citet{GilbankBalogh2008} are shown as blue triangles. It can be seen that clusters at $z\sim1.15$ have consistent faint-to-luminous ratios as in the field.  The ratios in clusters evolve strongly with redshift, while the ratios in the field show a much milder redshift dependence.}
 %For completeness we also show the ratios with no evolution correction applied (light brown squares) and with a $z_f = 3$ SSP evolution correction applied similar to the clusters (grey diamonds).} 
  \label{fig_dgr}
\end{figure*}

Another analogous way to probe the redshift evolution of the build up of the red sequence is to compare the faint-to-luminous ratio of clusters in different redshifts.  Similar to the faint end slope of the red sequence LF, various works have debated whether there is an evolution of the faint-to-luminous ratio with redshift \citep[see, e.g.][and references therein]{Capozzietal2010, Bildfelletal2012, Ceruloetal2016}.  The left panel of Figure~\ref{fig_dgr} compares the faint-to-luminous ratio of our sample with results from the literature \citep[][]{DeLuciaetal2007, Stottetal2007, Andreonetal2008, Stottetal2009, Capozzietal2010, Bildfelletal2012}.  To avoid biases due to inconsistent definitions, we only plot values from works that define the faint-to-luminous ratio in a way that is compatible with \citet{DeLuciaetal2007} in Figure~\ref{fig_dgr}.  Studies that use a different definition due to insufficient depth of the data \citep[e.g.][]{Gilbanketal2008, Ceruloetal2016} are not included in the comparison.  In most cases we use the ensemble-averaged ratio quoted in their works directly, except we have binned the \citet{Bildfelletal2012} results in redshift bins of $\Delta z = 0.1$.  Note that here we do not make any cut in halo mass or other cluster properties and simply include all values that are available.

It is evident from Figure~\ref{fig_dgr} that there is a general trend of decreasing faint-to-luminous ratio with increasing redshift when we compare our results at $z\sim1.15$ with those at lower redshifts.  It is worth noting that similar to the LFs, we also see a decrease in faint-to-luminous ratio in the GOGREEN sample compared to the ratio of the EDisCS clusters \citep[derived by][]{DeLuciaetal2007}.  The cluster-average GOGREEN faint-to-luminous ratio (${0.78^{+0.19}_{-0.15}}$) is consistent with the expected evolution predicted by \citet{GilbankBalogh2008}, which assumes a functional form of $(1+z)^{-1.8 \pm 0.5}$ and is derived through fitting the measured faint-to-luminous ratios of $z \lesssim 0.9$ cluster samples from the literature.  Our result is also consistent with the empirical relation predicted by \citet{Bildfelletal2012}, which has a parametrisation of $((0.88 \pm 0.15) z + (0.44 \pm 0.03))$ and is derived from 97 clusters at a redshift range of $0.05 \leq z \leq 0.55$.  Although the \citet{GilbankBalogh2008} and \citet{Bildfelletal2012} relations differ by a considerable amount at low redshift (which \citet{Bildfelletal2012} suggests is due to the way the photometry is measured), they give very similar predictions at $z\sim1$, hence we are not able to distinguish between the two relations.

The evolution of the faint-to-luminous ratio is simply a change in the proportion between the faint and bright red sequence population.  One caveat of using this quantity, as suggested by \citet{Crawfordetal2009}, is that it is difficult to ascertain whether the measured evolution is due to the bright end or the faint end of the LF, as a changing $M^{*}$ can also contribute to the observed evolution in the faint-to-luminous ratio.  Nevertheless, in Section~\ref{subsec:The total red sequence luminosity} we show that the bright end of the red sequence has been mostly in place since $z\sim1$ after we apply the passive evolution correction.  Combining this with the fact that the EDisCS LFs also show a similar bright end to that of the SDSS sample \citepalias[as seen in][]{Rudnicketal2009}, we can conclude that the evolution in faint-to-luminous ratio since $z\sim1.15$ is a result of the gradual build up of the faint end of the red sequence. This conclusion is consistent with previous works studying the cluster red sequence at lower redshifts \citep[e.g.][]{DeLuciaetal2007, Rudnicketal2009, Stottetal2009, Bildfelletal2012, Martinetetal2015, Zhangetal2017, Sarronetal2018}.

Despite the clear evolutionary trend, we note that the ratios of our clusters appear to be lower than those of the three $z>1$ clusters in \citet{Andreonetal2008}, although there seems to be good agreement of their $z<1$ measurements with the literature. One possible explanation is that this is simply due to cluster-to-cluster variation, as noted in other studies \citep[e.g.][]{DeLuciaetal2007, Crawfordetal2009}.  The different color selection used to define the red sequence (\textit{HST}/F775W - F850LP color was used in \citet{Andreonetal2008}) %, which may be problematic for $z>1$ clusters as both bands are bluer than the 4000\AA ~break)
 and methods to measure the ratio may also play a role to the discrepancy. Our result here is also in contrast with \citet{Ceruloetal2016}, who found no evolution in faint-to-luminous ratios (luminous-to-faint ratio in their work) in the red sequence of nine high redshift clusters at $0.8 < z < 1.5$.  They reported that the ratios of the high redshift clusters (and the fitted $\alpha$ of the LF) are consistent with the WINGS clusters at $z\sim0$, albeit with a large scatter.  We note that they adopt a different definition of bright and faint galaxies, which is brighter than the one used in \citet{DeLuciaetal2007} due to the insufficient depth of their data.  Using the same definition as in \citet{Ceruloetal2016} we find a faint-to-luminous ratio (${0.81^{+0.12}_{-0.11}}$), which is consistent with the one using the \citet{DeLuciaetal2007} definition.  This value, after converting to luminous-to-faint ratio, is consistent within $1\sigma$ with three out of six of $z>1$ clusters in the \citet{Ceruloetal2016} sample.   It is possible that the apparent discrepancy is due to the scatter in their high redshift sample, or due to the definition of faint galaxies (i.e. which can suppress the faint-to-luminous ratio in their low redshift comparison point), as the majority of the observed evolution comes from the very faint end of the red sequence.

%==== DGR - z =========================                                              %CHECKED_REFV2
%=====================================

%%%%%% Comparison to the field %%%%%%%
\subsection{Comparison to the field}                                                                  %CHECKED_REFV2
\label{sec:Comparison to the field}
Having established the gradual build up of the faint end of the cluster red sequence over time, we then investigate the environment dependence of this build up by comparing the faint-to-luminous ratio of the clusters to the field.  \citet{GilbankBalogh2008} performed a similar analysis and compared the cluster faint-to-luminous ratios up to $z\sim0.9$ to a field sample combining various surveys, including the SDSS \citep{Belletal2003, Baldryetal2004}, COMBO-17 \citep{Belletal2004}, the Millennium Galaxy Catalogue \citep{Driveretal2006}, COSMOS \citep{Scarlataetal2007} and VVDS \citep{Zuccaetal2006}.  In this work, we make use of the new COSMOS/UltraVISTA DR3 catalogue (Muzzin et al. in prep.) to construct the faint-to-luminous ratio of the field up to the redshift of the GOGREEN sample.  The UltraVISTA DR3 catalogue comprises $\sim50$-band optical and infrared photometry and is significantly deeper than the DR1 release.  The catalogue covers an area of $\sim$0.7 deg$^{2}$.\footnote{Note that we use the UltraVISTA DR1 catalogue instead for statistical subtraction of the EDisCS sample as it covers a larger area.}

We derive the faint-to-luminous ratios for the field in redshift bins of $\Delta z = 0.2$ using the spectroscopic and photometric redshift information, as well as the rest-frame $V$-band magnitudes derived using \textsc{EAZY} \citep{Brammeretal2008} in the UltraVISTA DR3 catalogue. To select passive galaxies we utilise the $UVJ$ classification used in \citet{Muzzinetal2013c}.  Although this is not the same color selection used for the GOGREEN clusters, we stress that the possible systematics resulting from the selection difference will not change the conclusion of the comparison (see Section~\ref{subsec:Contamination from dusty star forming galaxies} for a discussion).  We adopt the same definition of the faint-to-luminous ratio as in \citet{DeLuciaetal2004} for the field.  To convert the galaxy magnitudes to $z=0$, we compute an evolution correction for individual galaxies using their stellar population history parameters derived with FAST \citep{Krieketal2009} in the DR3 catalogue.  Using a single formation redshift evolution correction for all the field galaxies is not preferred as the color variation among passive galaxies in the field at a given redshift is much larger than the red sequence in clusters.  For each galaxy, we construct a model using the exponential decaying SFH parameters (age, $Z$, $A_V$ and $\tau$) and passively evolve this model down to $z=0$ to obtain the correction.  Various assumptions in deriving this correction have been tested, such as 1) keeping the same dust extinction at $z=0$ versus no more dust at $z=0$ and 2) using a SSP model with the same age (i.e. formation redshift) instead of an exponential decaying SFH.  The systematics due to these assumptions are kept as a range of the faint-to-luminous ratios at each redshift bin.  Even with the DR3 catalogue the evolution corrected rest-frame $V$-band depth is not deep enough for computing the number of faint galaxies in the $z\sim0.95$ and $z\sim1.15$ bins ($\sim0.2$ and $\sim0.7$ mag brighter than the limit, respectively), primarily due to the large color variation in the field population.  Hence for these two bins we follow the method described in Section~\ref{subsec:Faint-to-luminous Ratio} to extrapolate the number of faint galaxies to compute the faint-to-luminous ratio.  We have also included the uncertainty from cosmic variance following the recipe in \citet{Mosteretal2011}.

The right panel of Figure~\ref{fig_dgr} compares the cluster faint-to-luminous ratio to the field.  The light green shaded regions correspond to the $1\sigma$ regions of the field faint-to-luminous ratios, taking into account the variation of the evolution correction from different assumptions.  %For completeness, we also show the field faint-to-luminous ratio ratios without any evolution correction applied, and the ratios with an $z_f =3$ SSP evolution correction applied (the same one used for the clusters).  
From Figure~\ref{fig_dgr} it is evident that clusters have comparable faint-to-luminous ratios as in the field at $z\sim1.15$.  This is consistent with the results of \citet{vanderBurgetal2013}, who found that the shape of the stellar mass function of passive galaxies in clusters is comparable to the field at $z\sim1$.  At lower redshifts, the ratio in clusters becomes higher than that of the field.  Interestingly, while the faint-to-luminous ratios in clusters evolve strongly with redshift, the ratios in the field show a much milder redshift dependence and is consistent with no evolution.  This result is in contrast with the evolution seen by \citet{GilbankBalogh2008}.  We notice that, however, given the large uncertainties in the data points in \citet{GilbankBalogh2008} (see Figure~\ref{fig_dgr}), the only difference that is $>1\sigma$ is their highest redshift bin at $z\sim0.9$, which comes from first epoch VVDS data \citep{Zuccaetal2006}.  We therefore suspect the difference is mainly due to the depth and area coverage of the catalogue.

Figure~\ref{fig_dgr} demonstrates that the environment plays an important role in shaping the build up of the passive population.   The mild redshift dependence in the field ratios suggests that newly quenched galaxies are being added in both the bright end and the faint end of the population by a similar fraction.  Indeed, the number density of the bright and faint galaxies in the field have both grown by a factor of $\sim4$ from the highest redshift bin ($z\sim1.15$) to the lowest redshift bin ($z\sim0.15$).  Note that the split magnitude between bright and faint galaxies ($M_{V, \rm{vega}}= -20$) roughly corresponds to $\log(M_{*}/M_{\odot}) \sim10.2-10.5$\footnote{This $M/L$ conversion is derived using a range of SSP models, assuming $z_f =0.5-4.0$ and $Z=Z_{\odot}$.}. The mild redshift dependence in the field ratios is therefore consistent with previous studies on stellar mass function of passive galaxies in the field, which have found significant growth in both the high-mass end and low-mass end from $z\sim1$ to $z\sim0$ \citep[e.g.][]{Muzzinetal2013b, Tomczaketal2014}.  The growth is commonly attributed to various mass-quenching processes internal to the galaxies, such as energetic feedback from supernovae and stellar winds for low-mass galaxies \citep[e.g.][]{DekelSilk1986, Hopkinsetal2014} and ejective feedback from active galactic nuclei (AGN) for more massive galaxies \citep[e.g.][]{Boweretal2006, Terrazasetal2016}.  Given that the field ratios do not evolve strongly with redshift, it is evident that the strong redshift dependence of the cluster ratios is a result of the high-density environment and that preferentially low (or moderate) mass galaxies are quenched by the environment.  Additionally, the difference between the cluster and field ratios suggests that the quenching effects induced by environment are clearer at low redshifts.  This is consistent with the mostly mass-independent environmental quenching scenario established at low redshift \citep[e.g.][]{Pengetal2010}. Since the LF of star-forming galaxies is steeper at the faint end, a population of environmentally quenched galaxies would have a relatively high faint-to-luminous ratio. Recent studies of infalling galaxies in local groups and clusters have demonstrated that the ram pressure stripping of the cold gas in the galaxies when it passes through the ICM is the dominant mechanism \citep[e.g.][]{Bosellietal2016, Fillinghametal2016, Fossatietal2018}, although strangulation may also play a significant role \citep[for galaxies with $\log(M_{*}/M_{\odot}) < 11$, see e.g.][]{Fillinghametal2015, Pengetal2015}.

Although we will not be able to identify the main quenching mechanism at $z\sim1.15$ using only the faint-to-luminous ratios, from Figure~\ref{fig_dgr} we can see that the effect of environmental quenching in the ratios only emerges after $z<1$.  The naive interpretation is that environmental quenching is negligible, but this cannot be true as various works have shown that the passive fraction for massive galaxies in clusters are at least $30\%$ higher than the field at this redshift range, suggesting that environmental quenching is at work \citep[i.e. with a non-zero environmental quenching efficiency,][]{Nantaisetal2016, Nantaisetal2017, Foltzetal2018}.  The next simplest explanation is then that clusters at $z\sim1.15$ have higher fraction of both bright and faint passive galaxies than the field due to environmental quenching effects, but in a way that the faint-to-luminous ratio has to remain consistent with the field.

The implication of the above explanation is worth exploring, as it provides constraints on the relative fraction of bright and faint galaxies that are environmentally quenched.  To further investigate its connection to the environmental quenching mechanism, a robust measurement of the star forming galaxy luminosity function is required.  With the full GOGREEN sample, we will be able to study and model in detail the evolution of both passive and star-forming populations in these high-density environments up to $z\sim1.5$, but it is not possible with this preliminary analysis.

%%%%%%%%%%%%%%%%%%%%%%%%%%%%%%%%%%%%%%%%%%%
%%%%%%%%%     Summary      %%%%%%%%%%%%%%%%%%%%%%%%%
%%%%%%%%%%%%%%%%%%%%%%%%%%%%%%%%%%%%%%%%%%%

\section{Summary and Conclusions}                                                                  %CHECKED_REFV2
\label{sec:Conclusion}

In this paper we have presented the rest-frame $H$-band luminosity function and faint-to-luminous ratio of red sequence galaxies in seven clusters at $1.0 < z < 1.3$ from the Gemini Observations of Galaxies in Rich Early Environments Survey (GOGREEN).  We compare the composite red sequence LFs of these clusters with a sample of EDisCS clusters at $z\sim0.6$ to investigate the build up of the red sequence.  Our results can be summarised as follows:
  
\begin{itemize}
   \item The red sequence LF of all seven clusters shows a gradual decrease towards the faint end. By stacking the entire sample, we derive a shallow cluster-average faint end slope of ${\alpha \sim -0.35^{+0.15}_{-0.15}}$ and ${M_{H}^{*} \sim -23.52^{+0.15}_{-0.17}}$ using the MLE approach.
   \item We compare the composite LF of the GOGREEN clusters to a sample of EDisCS clusters at $z\sim0.6$.  After applying the passive evolution correction and renormalising the composite LF to have the same halo mass as the EDisCS clusters, we find that the bright end of the two LFs are consistent with each other.  This suggests that most of the bright passive population in clusters are already assembled at $z\sim1.15$.  The evolution corrected $M_{H}^{*}$ of the two LFs are also consistent with each other.  On the other hand, the composite LF of the GOGREEN clusters shows a shallower slope on the faint end compared to the EDisCS clusters (full EDisCS sample: $\alpha \sim -0.90^{+0.13}_{-0.15}$, selected EDisCS sample that are plausible descendants of GOGREEN clusters: $\alpha \sim -0.91^{+0.15}_{-0.17}$), implying that there is a build up of faint red sequence galaxies over time.
   \item By integrating the red sequence LFs of the passive evolution corrected and renormalised GOGREEN and EDisCS sample, we find that the total red sequence luminosities grow by $\sim31\% \pm 30\%$ (full EDisCS sample, $\sim13\% \pm 28\%$ if comparing with the selected EDisCS sample) over the $\sim2.6$ Gyrs between $z \sim 1.15$ and $z \sim 0.60$.  The growth comes mostly from the faint end ($M_H > -23$, passive evolution corrected), the mean luminosity of the faint end grows by almost a factor of two, while the bright end ($M_H \leq -23$) of the GOGREEN clusters is completely consistent with EDisCS.  Note that BCGs are excluded from the red sequence LFs of both samples.
    \item There is a general trend of decreasing faint-to-luminous ratio with increasing redshift when comparing the ratio of the GOGREEN clusters (${0.78^{+0.19}_{-0.15}}$) with literature values at lower redshifts, suggesting that the build-up of the faint red sequence galaxies occurs gradually since $z \sim 1.15$.  The amount of decrease is consistent with the evolution predicted in previous studies.
    \item Comparing the faint-to-luminous ratios of the clusters to those of the field, we find that they show different evolution with redshift.  At $z\sim1.15$, clusters have consistent faint-to-luminous ratios as the field.  The ratios of the field only show a mild redshift dependence since $z\sim1$.  We have explored various assumptions in deriving the passive evolution correction for the ratios of the field, therefore our findings are robust to reasonable uncertainties in the evolution corrections. The strong redshift evolution of the cluster ratios demonstrates that the environment plays an important role in shaping the build up of the passive population.  The fact that clusters show consistent faint-to-luminous ratios as in the field suggest that both bright and faint cluster galaxies experience the quenching effect induced by environment, and provide constraints on the relative fraction of bright and faint galaxies that are environmentally quenched at high redshift.
\end{itemize}

\acknowledgments

%\section*{Acknowledgements}
% Entry for the table of contents, for this guide only
%\addcontentsline{toc}{section}{Acknowledgements}
The authors would like to thank the anonymous referee for providing useful suggestions and comments. We also thank Ian Smail for providing suggestions that greatly improved the quality of the manuscript.  We thank Emiliano Munari for providing us with the \texttt{C.L.U.M.P.S.} code in advance of publication. This work is supported by the National Science Foundation through grants AST-1517863, and in part by AST-1518257 and AST-1815475.  This work is also supported by HST program numbers GO-13677/14327.01 and GO-15294,  and by grant number 80NSSC17K0019 issued through the NASA Astrophysics Data Analysis Program (ADAP). Support for program numbers GO-13677/14327.01 and GO-15294 was provided by NASA through a grant from the Space Telescope Science Institute, which is operated by the Association of Universities for Research in Astronomy, Incorporated, under NASA contract NAS5-26555.  Additional support was also provided by NASA through grants AR-13242 and AR-14289 from the Space Telescope Science Institute. JN is funded by the Universidad Andres Bello internal grant no. DI-18-17/RG.  PC acknowledges the support provided by FONDECYT postdoctoral research grant no 3160375.  RD gratefully acknowledges the support provided by the BASAL Center for Astrophysics and Associated Technologies (CATA) grant AFB-170002.

This paper includes data gathered with the Gemini Observatory, which is operated by the Association of Universities for Research in Astronomy, Inc., under a cooperative agreement with the NSF on behalf of the Gemini partnership: the National Science Foundation (United States), the National Research Council (Canada), CONICYT (Chile), Ministerio de Ciencia, Tecnolog\'{i}a e Innovaci\'{o}n Productiva (Argentina), and Minist\'{e}rio da Ci\^{e}ncia, Tecnologia e Inova\c{c}\~{a}o (Brazil). This work is based [in part] on observations and archival data obtained with the Spitzer Space Telescope, which is operated by the Jet Propulsion Laboratory, California Institute of Technology under a contract with NASA. Support for this work was provided by an award issued by JPL/Caltech.\\
%\bsp
%%%%%%%%%%%%%%%%%%%%%%%%%%%%%%%%%%%%%%%%%%%%%%
%IRS acknowledges support from the ERC Advanced Grant DUSTGAL (321334) and STFC (ST/P000541/1).

%%%%%%%%%%%%%%%%%%%% REFERENCES %%%%%%%%%%%%%%%%%%
% The best way to enter references is to use BibTeX:

\bibliographystyle{aasjournal}
\bibliography{gg_lf_6_apj_arxivsubmission} % if your bibtex file is called example.bib

%%%%%%%%%%%%%%%%%%%%%%%%%%%%%%%%%%%%%%%%%%%%
%%%%%%%%%%%%%            APPENDIX              %%%%%%%%%%%%%%%%%    (ALL CHECKED_REFV2)
%%%%%%%%%%%%%%%%%%%%%%%%%%%%%%%%%%%%%%%%%%%%
\appendix

\section{Matching of filter passbands for statistical background subtraction}                   %APP A CHECKED_REFV2
\label{Matching of filter passbands for statistical background subtraction}
As we mentioned in Section~\ref{subsec:Cluster membership}, in the ideal case of statistical field subtraction the field catalogue should have the same depth and passband as the cluster photometry.  Although this is often not the case in reality, occasionally one can utilise the additional passbands available in the control field catalogue to achieve a nearly perfect match.  In this section we expand on the discussion in Section~\ref{subsubsec:Control field catalogue} on how we use both the $z$ and $y$-band data of HSC-SSP to match the passband of the GMOS $z'$-band.

\begin{figure}
  \centering
  \includegraphics[scale=0.55]{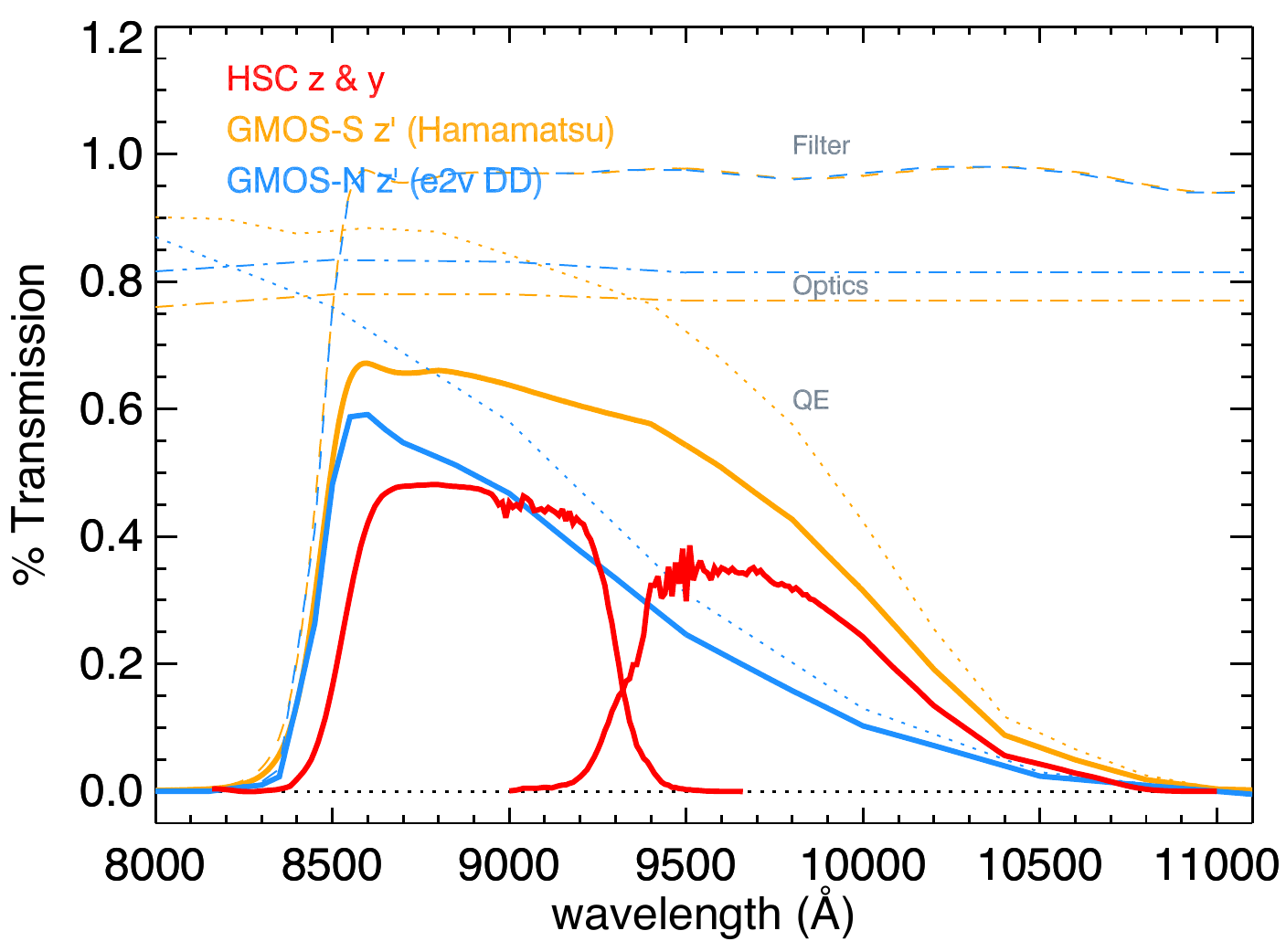}
  \caption[Filter Transmission]{The wavelength response of the Gemini/GMOS $z'$ filters and the Subaru/HSC $z $ and $y$ filters.  Orange lines show the total wavelength response (solid) of the GMOS-S $z'$ filter (dashed), the Gemini South telescope response (dot-dashed) and the QE of the Hamamatsu detector (dotted).  Similarly, blue lines correspond to the GMOS-N $z'$ filter (dashed), the Gemini North telescope response (dot-dashed) and the QE of the e2V DD detector (dotted).  Red lines correspond to the HSC total response with $z$ and $y$ filters.  It can be seen that the GMOS $z'$-band has a much longer effective wavelength compared to the HSC $z$-band.  Hence, a combination of the HSC $z $ and $y$ filters is used to match the GMOS $z'$ filters for statistical background subtraction.}
  \label{fig_filter_qe}
\end{figure}

Figure~\ref{fig_filter_qe} shows the transmission curve of the Gemini/GMOS $z'$ filters and Subaru/HSC $z$ and $y$ filters used in this work.  The wavelength coverage of the GMOS-S and GMOS-N $z'$ filters are much wider and extend further to the red compared to the HSC $z$ filters. In fact, the GMOS $z'$ filters cover further to the red wavelength compared to most $z$-band filters, due to the outstanding red-sensitivities of the e2v DD and Hamamatsu detectors on Gemini North and South (blue and orange dotted curve).  It can be seen that the GMOS $z'$ band almost covers both HSC $z$ and $y$-bands.

%==== Filter transmission ==================                                                          %CHECKED_REFV2
%=====================================

\begin{figure}
  \centering
  \includegraphics[scale=0.65]{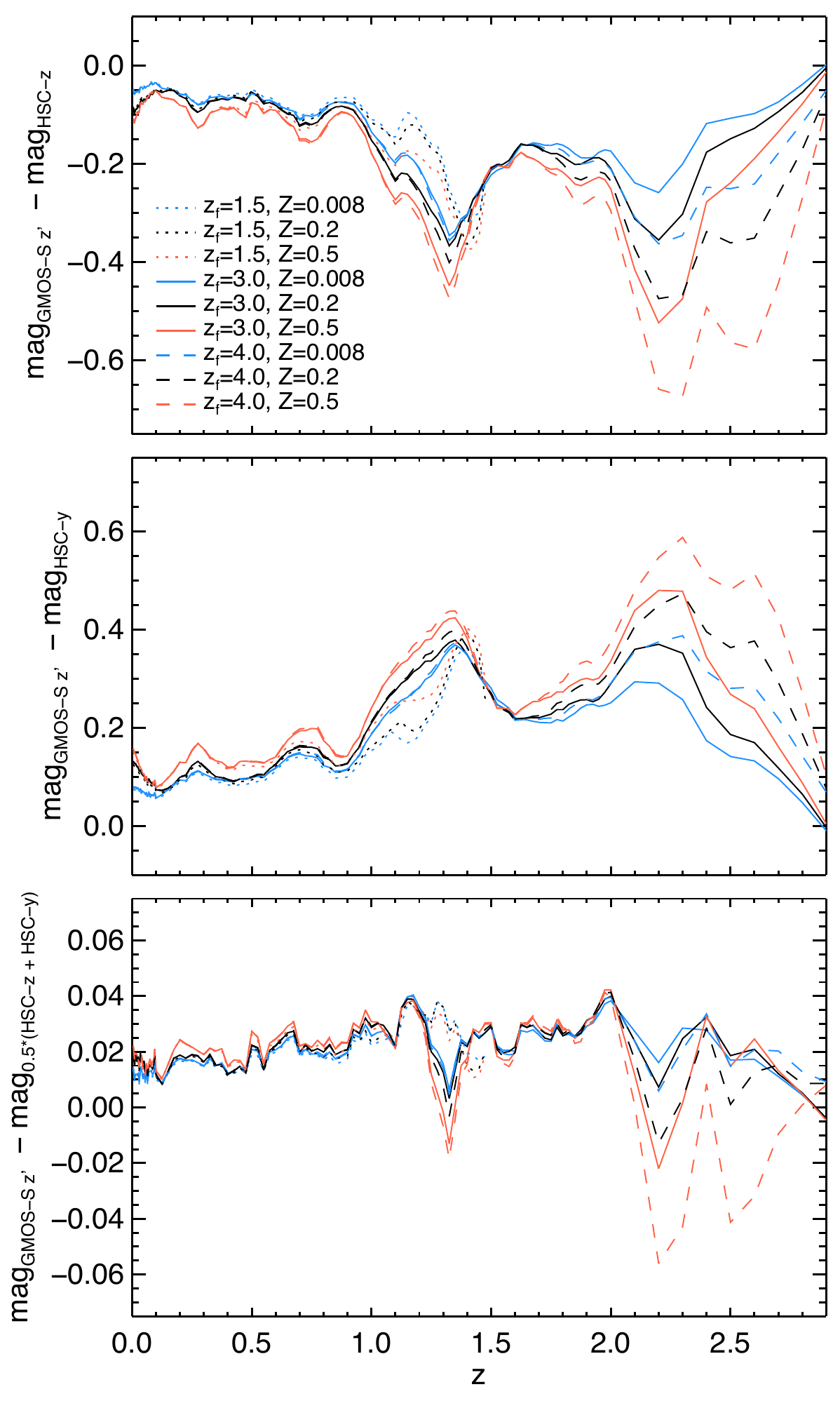}
  \caption[Filter Transmission]{The color term between the GMOS-S $z'$ and HSC filters with redshift. Top to bottom: Difference in apparent magnitudes of SSP models between GMOS $z'$ and HSC $z$, GMOS $z'$ and HSC $y$, and GMOS $z'$ and a combination of HSC $z$ and $y$ as a function redshift.  Blue, black and red curves correspond to models with a metallicity of $Z=0.4Z_\odot, Z_\odot, 2.5Z_\odot$, while the solid, dotted and dashed line styles correspond to models with formation redshift of $z_f = 1.5,3.0,4.0$, respectively.}
\label{fig_filter_colorterm}
\end{figure}

The next step is to derive the color term between the GMOS $z'$ and HSC filters.  We test the stability of the color terms with redshift by computing apparent magnitudes of BC03 SSP models with a range of formation redshifts and metallicities in different filters.  The result of GMOS-S is shown in Figure~\ref{fig_filter_colorterm} as an example.  The optimal color term between two filters for statistical field subtraction should be flat at all redshifts, so that one does not under- or over-subtract galaxy populations at a certain redshift range and bias the resulting LF.  Nevertheless, due to the difference in the wavelength coverage between the GMOS and HSC filters,  one can see from the top and middle panels of Figure~\ref{fig_filter_colorterm} that there are various troughs and peaks at different redshifts, due to spectral features being redshifted into the filter coverage.  Hence, we use a combination of HSC $z$ and $y$-band magnitudes to match the passband of the GMOS $z'$ filters, shown in the bottom panel. It is clear that the stability of the color term with redshift is vastly improved when a combination of the two bands is used.  The variation ($\sim 0.02$~mag) is smaller than the photometric uncertainty of most objects we are interested in.  The color term used for GMOS-S is $\sim0.02$ mag, while the one used for GMOS-N is $\sim0.01$ mag.  We note that using other SSP models \citep[e.g.][]{Maraston2005} gives similar results as these are broad-band colors.  These color terms are then applied while comparing the color-magnitude diagram of the cluster and the field to obtain cluster membership probabilities for individual galaxies. %(Section~\ref{subsubsec:Membership probabilities}).

% Through comparing the catalogues in color-magnitude space, the excess in number counts can be converted into a probability of being a cluster member.  This technique allows us to make full use of the deep GOGREEN $z'$ and $[3.6]$ photometry, at the same time without needing to derive photometric redshifts.
%==== Filter comparison ===================                                                          %CHECKED_REFV2
%=====================================

%\section{Comparison of the spectroscopic membership with photometric membership probabilities}    %APP B REMOVED, see ver 3.

%%%%%%%%%%%% Halo mass section moved to Appendix
\section{Does the LF depend on halo mass?}                                                            %APP C CHECKED_REFV2
\label{app:Does the LF Depend on Halo mass}
One interesting question regarding cluster LFs concerns the possible dependence of the red sequence LF on cluster mass.  Previous studies at intermediate redshifts ($0.5 < z < 0.8$) have shown, by examining faint-to-luminous ratio or $\alpha$ of the LF, that high mass clusters may evolve at a faster rate than low mass ones at similar redshifts \citep[e.g.][]{Tanakaetal2005, DeLuciaetal2007, Gilbanketal2008, Martinetetal2015}.  In this section we examine this dependence with our sample at $z\sim1$.

We split our sample into two bins in halo mass (see Table~\ref{tab_data_summary}) and derive a composite LF for each bin.  The low mass bin comprises five clusters with $13.9 \leq \log (M_{200} / M_{\odot}) \leq 14.6$, while the high mass bin comprises the remaining two massive clusters with $14.8 \leq \log (M_{200} / M_{\odot}) \leq 14.9$.  We find that $\alpha$ and $M_{H}^{*}$ of the two composite LFs are consistent with each other.  Using a radius limit of $R \leq 0.5 R_{200}$ instead of $R \leq 0.75$ Mpc also gives a similar result.  The best fitting $\alpha$ and $M_{H}^{*}$ can be found in Table~\ref{tab_fit_summary}. Similarly, the faint-to-luminous ratios show no obvious trend with halo mass in our sample.  %From v528, v535

However, it is entirely possible that the dependence is hindered by the small number of clusters here and the large uncertainties in the derived quantities.  \citet{Gilbanketal2008} measured a $\sim2\sigma$ difference in the faint-to-luminous ratios (luminous-to-faint ratio in their work) between poor and rich clusters at $z\sim0.9$ with $98$ clusters.  Therefore it is very likely that with a factor of 10 smaller sample we are not able to see such a small difference.  With the full GOGREEN sample which triples the number of clusters used here, we might be able to discern and constrain the possible dependence with halo mass.

%%%%%%%%%%% Radius section moved to Appendix
\section{Is there any radial dependence?}                                                                 %APP D CHECKED_REFV2
\label{app:Is there any radial dependence}
A number of works on local to intermediate redshift clusters have reported that the shape of the LF varies with radius, they found that the faint end slope $\alpha$ becomes less positive (i.e. steeper) with increasing selection radius \citep[e.g.][]{Popessoetal2006, Barkhouseetal2007, Crawfordetal2009}. Other works, however, find little or no evidence of radial dependence \citep[e.g.][]{Barrenaetal2012, Martinetetal2015, Martinetetal2017}, or that the difference can only be seen in the densest region of the cluster \citep[$R<0.25 R_{200}$,][]{Annunziatellaetal2014, Annunziatellaetal2016}.

\begin{figure}
  \centering
  \includegraphics[scale=0.515]{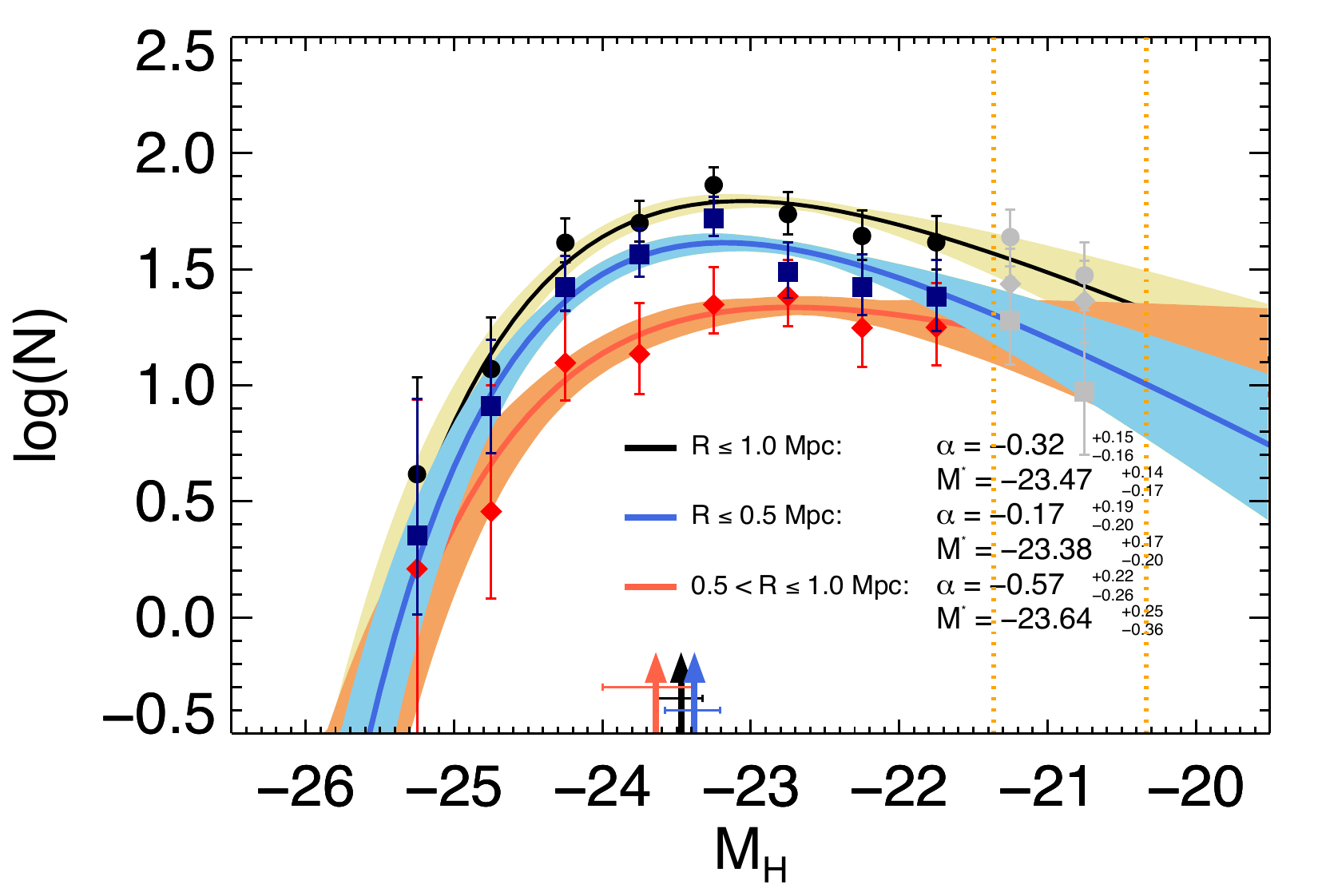}
   \caption{Cluster-centric radial dependence of the rest-frame $H$-band red sequence LF of the seven GOGREEN clusters.  Same as Figure~\ref{fig_lf_stack_all}, but for $R \leq 1.0 $ Mpc.  The blue and red lines correspond to the best-fitting Schechter functions to the inner and outer LFs with a selected radial range of $R \leq 0.5$ Mpc and $0.5  < R \leq 1.0$ Mpc, respectively.}
  \label{fig_lf_rad_depend}
\end{figure}

Here we investigate the radial dependence of the LF in our sample.  Due to the small number of clusters we have, we split the LF that is derived with a selection radius of $R \leq 1.0$ Mpc into two radial sections: $R \leq 0.5$ Mpc (inner) and $0.5  < R \leq 1.0$ Mpc (outer).  The result is shown in Figure~\ref{fig_lf_rad_depend}.  We found that the outer radial section shows a steeper slope with $\alpha = -0.57^{+0.22}_{-0.26}$ than the inner one ($\alpha = -0.17^{+0.19}_{-0.20}$), although the difference is only $\sim1\sigma$ given the large uncertainties in the derived $\alpha$. Comparing the number counts, the $R \leq 1.0$ Mpc LF is mostly dominated by the inner section. Unfortunately we cannot test if this difference is due to the cluster core as suggested by \citet{Annunziatellaetal2014,Annunziatellaetal2016}, as there is not enough number statistics to accurately determine Schechter parameters if we adopt a selection radius of $R<0.25 R_{200}$.  Similar to the halo-mass dependence, we will be able to better constrain this radial dependence with the full GOGREEN sample.\\\\\\

%==== Radial Dependence ========================                                    %CHECKED_REFV2
%Black circles are derived using actual numbers of cluster galaxies, while the red circle is computed using extrapolation of the fitted LF due to insufficient depth of the image of SPT0205.
%==========================================

%%%%%%%%%%% Effect of blending
\section{Potential effect of source blending on the LF}                                             %APP E  CHECKED_REFV2
\label{app:Potential effect of source blending on the LF}
In this work we use PSF-matched $2''$ diameter $z'-[3.6]$ color measurements and pseudo-total $3.6~\mu$m magnitudes to construct the GOGREEN LF (see Section~\ref{subsec:Source detection and PSF-matched Photometry} for details).  A common concern with IRAC photometry is source blending issues due to its large-FWHM PSF. Source blending can result in inaccurate photometry (i.e. flux contamination from neighbouring objects) or the failure to detect a galaxy entirely (i.e. missing detections).  In the case of our analysis, missing galaxies because of blending is not a concern as we used unconvolved $z'$-band images instead of the $3.6~\mu$m images as the detection band.  Inaccurate photometry due to source blending, on the other hand, can be a source of uncertainty or bias to our LF. Here we give an estimate of how flux contamination from neighbors would affect our results.

Inaccurate photometry from source blending mainly affects the selection of red sequence galaxies in color-magnitude space.  One thing to note is that the effect of source blending can go in two ways: 1) objects that should lie outside the red sequence selection region can scatter into the red sequence, thus increasing the apparent number of red sequence galaxies. 2) Objects in the red sequence can also drop out of our red sequence selection due to flux contamination from their neighbours, and thus decreasing the apparent number of red sequence galaxies.  Since the effect of blending is usually more severe for faint objects, the former would result in a LF that is steeper (i.e. $\alpha$ being less positive) and the latter would result in a LF with an apparent $\alpha$ that is more positive. In this section we focus on the latter as its effect is more relevant to our findings.

The first step is to determine the conditions under which the photometry of the object of interest is contaminated by its neighbor.  To do this we have constructed a simulation by putting sets of objects and `neighbors' as functions of separation distance, $3.6~\mu$m object magnitude and neighbour magnitude on the images.  All sources are set to a size of $\sim1.7$ kpc, the median size of a passive galaxy with $\log(M / M_{\odot}) \sim 10.7$ at $z\sim1.15$ \citep{vanderWeletal2014}, and are convolved with the $3.6~\mu$m PSF.  By measuring the $2''$ diameter aperture magnitudes, we then define a region on the separation distances-object magnitudes-neighbor magnitudes plane where the resultant colors deviate more than 0.125 mag from the true value as `contaminated'. This choice of 0.125 mag will be explained below.  Objects that are within the region are considered to be severely affected by blending.  Using these regions as criteria we then pick out objects in our cluster photometric catalogues that have potentially problematic color due to source blending. A relation of the percentage of these contaminated objects as a function of magnitude is derived.  We find that the percentage of these objects increases with $3.6~\mu$m magnitudes; for the GOGREEN clusters on average only $\sim5\%$ of objects have potentially problematic color at $[3.6] = 18$, while at $[3.6] = 23$ $\sim25\%$ of objects may be contaminated according to the criteria.

The next step is to investigate how these objects may impact our results. In addition to the red sequence LF, we construct a stacked LF for the GOGREEN clusters using regions that are right above or below the red sequence (i.e. the two strips on the CMD that are above and below the red sequence selection region), following the same procedure described in Section~\ref{sec:Constructing the Luminosity Function}.  The width of these two regions are chosen to be $0.25$ mag to contain the majority of objects that can drop out of the red sequence due to source blending.  Although it may seem arbitrary to use such width, we stress that increasing the width does not change the conclusion of this section as it is (increasingly) less likely to have objects with larger magnitude deviation from source blending. The abovementioned criteria to select `contaminated' sources is therefore chosen to be half of the width of these regions (i.e. 0.125 mag) as we can assume that this is the average magnitude deviation the objects require to have to drop out of our red sequence selection.

\begin{figure}
  \centering
  \includegraphics[scale=0.517]{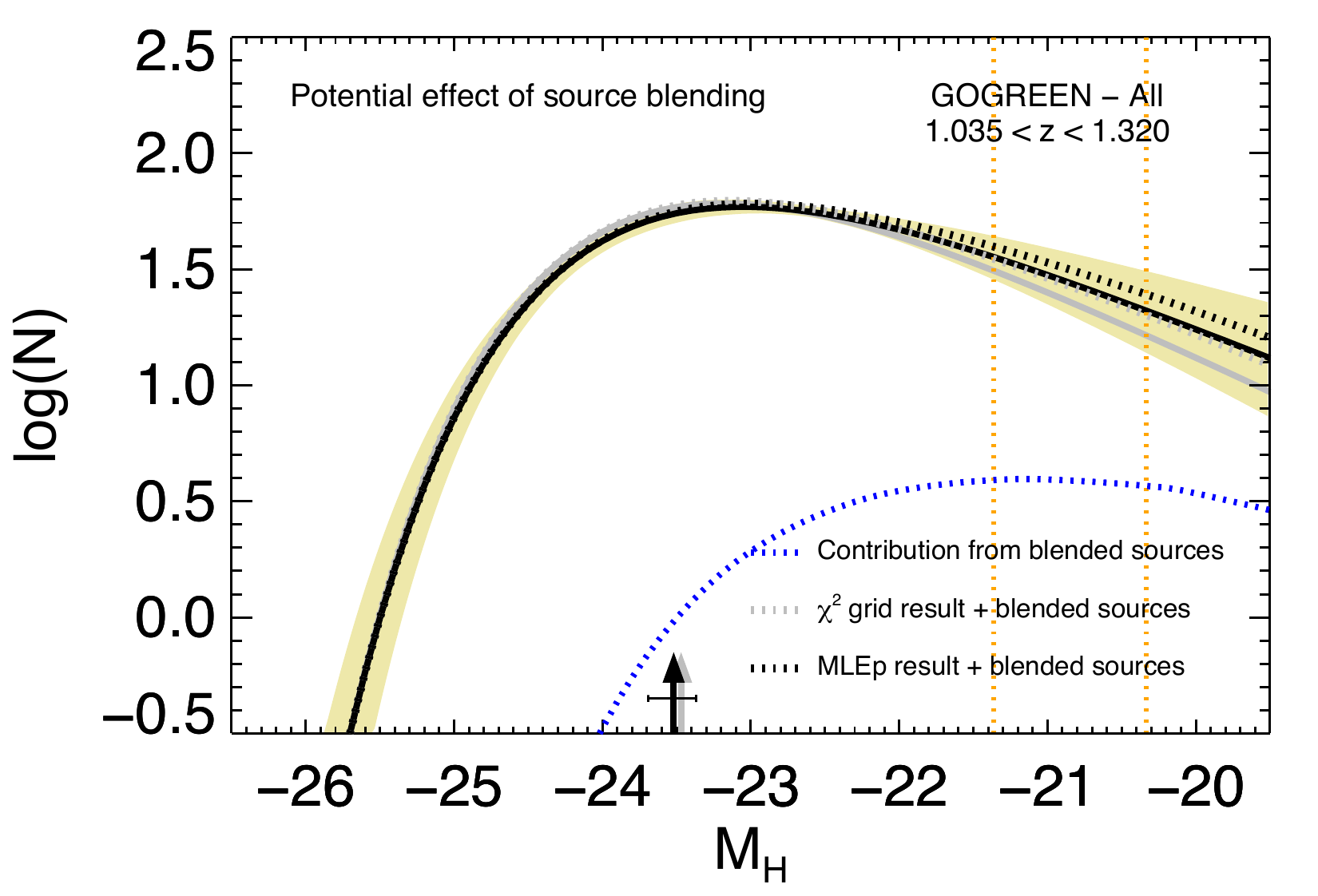}
  \caption{Potential effect of source blending to the red sequence LF of the seven GOGREEN clusters. The solid black and grey lines, as well as the vertical arrows correspond to the same LF fits as Figure~\ref{fig_lf_stack_all}.  The blue dotted line correspond to the potential contribution from blended sources that drop out of the red sequence and thus are not accounted for in the LF.  The black and grey dotted lines are the sum of the LF fits with this contribution from blended sources. It can be seen that the resultant LFs are still within the $1\sigma$ uncertainties of the LF estimated from the MLE method, therefore source blending is unlikely to affect our conclusions.}
  \label{fig_lf_stack_allandsb}
\end{figure}

We then combine the LF of these regions with the relation of the percentage of the contaminated objects as a function of magnitude to estimate the number of objects that drop out of the red sequence because of blending. The blue dotted line in Figure~\ref{fig_lf_stack_allandsb} shows the result.  As expected, there are more of these sources at the faint end compared to the bright end.  Even with the addition of these objects to our red sequence LF, we find that the resulting LFs (black and gray dotted lines) are still within $1\sigma$ uncertainties of our red sequence LF.  Therefore source blending is unlikely to affect our conclusions.  We also stress that the blue dotted line is a conservative estimate, as it only concerns sources that drop out of red sequence due to source blending.  As we mentioned above in reality there are also sources that scatter into the red sequence as well which affects the LF in the opposite way, thus the overall effect of source blending is even smaller.

%==== LF + SB - all ==============                                      %CHECKED_REFV2
\end{document}